\documentclass[10pt]{iopart}
\usepackage{iopams,setstack,epsfig,psfrag}
\usepackage[usenames, dvipsnames]{color}
\usepackage{rawfonts}
\input prepictex
\input pictex
\input postpictex

\def\@begintheorem#1#2{\par\bgroup{\sc #1\ #2. }\it\ignorespaces}
\def\@opargbegintheorem#1#2#3{\par\bgroup{\sc #1\ #2\ (#3). }\it\ignorespaces}
\def\@endtheorem{\egroup\par}

\eqnobysec

\usepackage{graphicx}
\usepackage{subfigure}
\usepackage{epstopdf}

\input prepictex
\input pictex
\input postpictex

\def\Pr{{\it Proof:  }}
\def\qed{$\Box$}
\def\L{\left(}   \def\R{\right)}

\def\C#1{{\cal{#1}}}

\def\Ref#1{(\ref{#1})}

\def\sfrac#1#2{\hbox{\normalsize $\frac{#1}{#2}$}}

\newtheorem{theo}{Theorem}
\newtheorem{lemm}{Lemma}

\begin{document}

\title{Polygons pulled from an adsorbing surface}
\author{A J Guttmann$^*$, E J Janse van Rensburg$^\dagger$, I Jensen$^*$ and S G  Whittington$^\ddagger$ }
\address{
{}$^*$School of Mathematics and Statistics, The University of Melbourne, Parkville, Victoria 3010, Australia \\
{}$^\dagger$Department of Mathematics \& Statistics, York University, M3J 1P3, Toronto, Canada \\
{}$^\ddagger$Department of Chemistry, University of Toronto, M5S 3H6, Toronto, Canada 
}
\begin{center}
 \today  \end{center}

\begin{abstract}
We consider self-avoiding lattice polygons, in the hypercubic
lattice, as a model of a ring polymer adsorbed at a surface and
either being desorbed by the action of a force, or pushed towards the surface.  
We show that, when there is no interaction with the surface, then 
the response of the polygon to the applied force is identical (in 
the thermodynamic limit) for two ways in which we apply the force.  
When the polygon is attracted to the surface then, when
the dimension is at least $3$, we have a complete characterization of the 
critical force--temperature curve in terms of the behaviour, (a) when there is 
no force, and, (b) when there is no surface interaction.   For the $2$-dimensional case
we have upper and lower bounds on the free energy.  
We use both Monte Carlo and exact enumeration and series analysis
methods to investigate the form of the phase diagram
in two dimensions.  We find evidence for the existence
of a \emph{mixed phase} where the free energy depends on
the strength of the interaction with the adsorbing line and
on the applied force. 
\end{abstract}

\submitto{J Phys A}
\maketitle

\maketitle
\section{Introduction}
The theory of polymer adsorption at an impenetrable surface is a well established subject.  
Useful reviews can be found in \cite{DeBell} and \cite{Rensburg2000}.  One of 
the standard models is self-avoiding walks, 
or SAWs, confined to a half-space and interacting
with the confining line or plane.  For this problem we have some rigorous 
results \cite{HTW1982,Rensburg1998} that establish the existence of a phase transition and provide
useful information about the behaviour of the free energy as the temperature
is varied.  More detailed information comes from a variety of numerical investigations
that, among other things, give quite precise information about the location of 
the phase transition \cite{Beaton2012,Guim1989,Guttmann2014,Hegger1994,Rensburg2004}
and strongly suggest that the transition is second order 
\cite{Guim1989,Hegger1994,Rensburg2004}.

With the advent of atomic force microscopy \cite{Haupt1999,Zhang2003} it has become
possible to pull an adsorbed polymer off a surface at which it is adsorbed.  In principle it is possible
to measure the temperature dependence of the critical force for desorption and the stress-strain
curves.  It is only quite recently that the effects of a force have been investigated 
for the self-avoiding walk model \cite{Beaton2015,Guttmann2014,Rensburg2009,JvRW2013,Krawczyk2005,Mishra2005,Binder2012}.

Consider the $d$-dimensional hypercubic lattice, ${\mathbb Z}^d$, and attach a coordinate system
$(x_1,x_2, \dots x_d)$ so that each vertex of the lattice has integer coordinates.  Suppose
that $c_n^+$ is the number of $n$-edge self-avoiding walks that start at $(0,0, \ldots 0)$ and have 
all vertices in the half-space $x_d \ge 0$.  
It is known  \cite{Whittington1975} that 
\begin{equation}
\lim_{n\to\infty} n^{-1} \log c_n^+ = \kappa_d
\end{equation}
where $\kappa_d$ is the \emph{connective constant} of the lattice \cite{HM54}.
Each vertex of the walk in the hyperplane
$x_d=0$ is called a \emph{visit}.  Let $c_n^+(v,h)$ be the number of these walks 
with $v+1$ visits and having the $x_d$-coordinate of their last vertex equal to 
$h$, which we call the \emph{height} of the last vertex.  Define the 
partition function 
\begin{equation}
C_n^+(a,y) = \sum_{v,h} c_n^+(v,h) a^v y^h.
\label{eqn:sawpf}
\end{equation}
It is known \cite{JvRW2013} that the limit 
\begin{equation}
\lim_{n\to\infty} n^{-1} \log C_n^+(a,y) = \psi(a,y)
\label{eqn:sawfreeenergy}
\end{equation}
exists.  $\psi (a,y)$ is the (reduced, limiting) free energy.

We can interpret the two fugacities $a$ and $y$ as
\begin{equation}
a = \exp (-\epsilon /k_BT) \quad \mbox{and} \quad y = \exp (f/k_BT)
\label{eqn:fugacities}
\end{equation}
where $k_B$ is Boltzmann's constant, $T$ is the absolute temperature, $\epsilon$
is the energy associated with a vertex in the surface and $f$ is the force applied at
the last vertex, normal to the surface.  For adsorption to occur $\epsilon < 0$ so that
the interaction with the surface is attractive.  For the walk to be desorbed
by the action of the force, $f > 0$.

If we set $f=0$, so that $y=1$, we have the pure adsorption problem.  Write 
$\psi(a,1) = \kappa (a)$, the free energy for pure adsorption.  
We know \cite{HTW1982,Rensburg1998}
that there is a critical value of $a$, $a_c > 1$, such that $\kappa (a) = \kappa (1)=\kappa_d$
for $a \le a_c$ and $\kappa (a) > \kappa (1)$ for $a > a_c$ so that the free energy is 
singular at $a=a_c$, corresponding to the adsorption transition.  Similarly if we 
set $\epsilon = 0$ so that $a=1$ we have no (attractive) interaction with the surface
and there is no adsorbed phase.  The free energy is $\lambda (y) = \psi (1,y)$
and we know that $\lambda (y)$ is singular at $y=y_c =1$ \cite{Beaton2015}.
There is a transition from a \emph{free phase} to a \emph{ballistic phase} at $y=1$
\cite{Beaton2015}, see also \cite{Ioffe_Velenik_2008,Ioffe_Velenik_2010}.

Returning to the full problem, we know \cite{JvRW2013} that 
for $a  \ge  a_c$ and $y \ge 1$
\begin{equation}
\psi(a,y) = \max [\kappa (a), \lambda (y)].
\label{eqn:sawferelation}
\end{equation}
This gives a complete characterization of the phase boundary between the 
adsorbed and ballistic phases in terms of the behaviour when $\epsilon = 0$ and 
when $f=0$.    This was used in \cite{Guttmann2014} to give very precise numerical estimates
of the location of the phase boundary when $d=2$.  Numerical estimates
using a different approach are given in \cite{Mishra2005} for both 
$d=2$ and $d=3$.  In addition we know
\cite{Guttmann2014} that the phase transition from the adsorbed to the ballistic
phase is first order.

%

These results raise a variety of new questions. For the self-avoiding walk model,
what happens if the force is applied somewhere other than at the last vertex? What
happens if the force is applied at an angle to the surface? Not all polymers are
linear and there are interesting questions about the behaviour of ring polymers or
branched polymers when they are pulled off a surface at which they are adsorbed.
That is, how does the architecture of the polymer affect its behaviour?

Does it matter where the 
force is applied?  If the force is applied at the \emph{top} vertex, \emph{i.e.} at the 
vertex furthest from the surface, this is equivalent to confining the walk  (or polygon, etc.) 
between two parallel lines or planes and requiring at least one vertex in each plane, 
then applying a force to move the confining plane.  For the walk problem
we know that the limiting free energy is the same when the force is applied in this way 
or at the terminal vertex \cite{RensburgWhittington2016a,RensburgWhittington2016b}. 
If the force is applied at an interior vertex, the free energy depends on which 
vertex the force is applied to, and in some circumstances, an additional phase can be present
\cite{RensburgWhittington2017}.  Some results for staircase polygons \cite{Beaton2017} 
suggest a phase diagram with a mixed adsorbed and ballistic phase.

In this paper we begin to investigate the issue of polymer 
architecture.  We consider a ring polymer 
adsorbed at a surface, being pulled off the surface by a force applied 
in a particular way.







\begin{figure}[t]
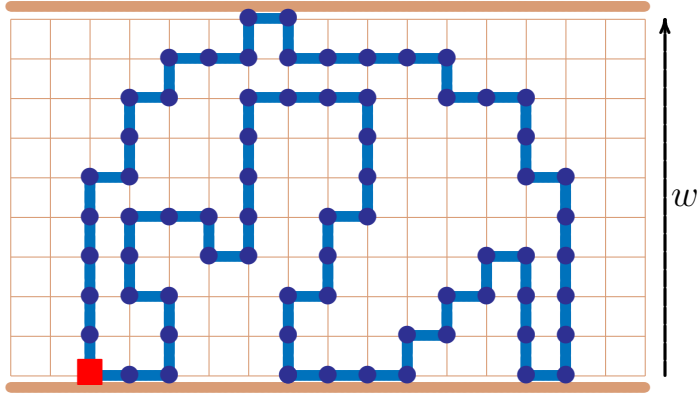

\beginpicture
\setcoordinatesystem units <1.5pt,1.5pt>
\setplotarea x from -140 to 110, y from -10 to 100
\setplotarea x from -80 to 80, y from 0 to 90

\color{Tan}
\grid 16 9 
\color{Tan}
\setplotsymbol ({$\bullet$})
\plot -80 93 80 93 /
\plot -80 -3 80 -3 /

\color{RoyalBlue}
\plot -60 0 -50 0 -40 0 -40 10 -40 20 -50 20 -50 30 -50 40 -40 40 -30 40 -30 30 
-20 30 -20 40 -20 50 -20 60 -20 70 -10 70 0 70 10 70 10 60 10 50 10 40 0 40 0 30 
0 20 -10 20 -10 10 -10 0 -10 0 0 0 10 0 20 0 20 10 30 10 30 20 40 20 40 30 50 30 50 20 50 10 
50 0 60 0 60 10 60 20 60 30 60 40 60 50 50 50 50 60 50 70 40 70 30 70 30 80 20 80
10 80 0 80 -10 80 -10 90 -20 90 -20 80 -30 80 -40 80 -40 70 -50 70 -50 60 -50 50 -60 50
-60 40 -60 30 -60 20 -60 10 -60 0   /
\color{Blue}
\multiput {\LARGE$\bullet$} at 
-60 0 -50 0 -40 0 -40 10 -40 20 -50 20 -50 30 -50 40 -40 40 -30 40 -30 30 
-20 30 -20 40 -20 50 -20 60 -20 70 -10 70 0 70 10 70 10 60 10 50 10 40 0 40 0 30 
0 20 -10 20 -10 10 -10 0 -10 0 0 0 10 0 20 0 20 10 30 10 30 20 40 20 40 30 50 30 50 20 50 10 
50 0 60 0 60 10 60 20 60 30 60 40 60 50 50 50 50 60 50 70 40 70 30 70 30 80 20 80
10 80 0 80 -10 80 -10 90 -20 90 -20 80 -30 80 -40 80 -40 70 -50 70 -50 60 -50 50 -60 50
-60 40 -60 30 -60 20 -60 10 -60 0   /

\color{red}
\multiput {\Large$\blacksquare$} at  -60 1 /

\color{black} \normalcolor

\setplotsymbol ({.})
\arrow <5pt> [.2,.67] from 85 0 to 85 90
\put {\Large$w$} at 90 45 
 
\endpicture
\label{fig1}

\caption{A polygon in a slab of width $w$.  If $w = \infty$ the polygon is in a half-space.  }
\end{figure}

\begin{figure}[h!]
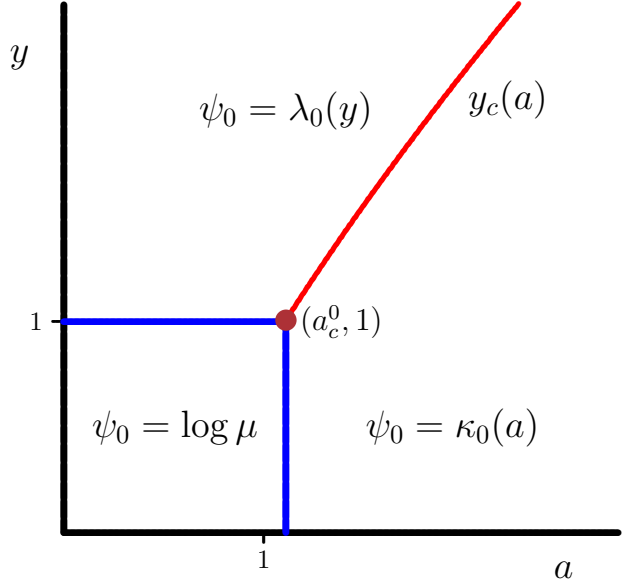

\beginpicture
\setcoordinatesystem units <2.1pt,2pt>
\setplotarea x from -40 to 100, y from -10 to 100

\color{black}
\setplotsymbol ({$\cdot$})
\plot -2 40 0 40 /   \plot 36 -2 36 0 /

\setsolid

\setplotsymbol ({\tiny$\bullet$})
\plot 0 100 0 0 100 0 /

\color{black}
\put {\Large$y$} at -8 90
\put {\Large$a$} at 90 -7
\put {$1$} at -5 40
\put {$1$} at 36 -5
\put {\Large$y_c(a)$} at 80 82
\put {\large$(a_c^0,1)$} at 50 40

\put {\Large\hbox{$\psi_0=\log \mu$}} at 20 20 
\put {\Large\hbox{$\psi_0=\lambda_0(y)$}} at 40 80 
\put {\Large\hbox{$\psi_0=\kappa_0(a)$}} at 70 20

\color{blue}
\plot 40 0 40 40 0 40 /

\setplotsymbol ({\LARGE$\cdot$})
\color{red}
\setquadratic
\plot 40 40 59 69 82 100 /
\setlinear
\color{Maroon}
\put {\huge$\bullet$} at 40 40

\color{black}
\normalcolor
\endpicture
\caption{The phase diagram of pulled adsorbing polygons in $d\geq 3$ dimensions.
For $y\leq 1$ and $a\leq a_c^0$ the free energy is $\psi_0 = \log \mu$.
This corresponds to a free phase with phase boundaries separating it
from the ballistic phase at $y=1$ and from the adsorbed phase.
In the ballistic phase the free energy is $\psi_0 = \lambda_0(y) = 
\lambda(\sqrt{y})$.  In the adsorbed phase the free energy is
$\psi_0 = \kappa_0(a) = \kappa(a)$.  The solution of 
$\lambda_0(y) = \kappa_0(a)$ is a critical curve $y_c(a)$ separating the ballistic
and adsorbed phases.  This phase boundary is first order.}
\label{figure22}
\end{figure}

\section{Definition of the model and statement of results}
\label{sec:definition}

A standard model of ring polymers is \emph{self-avoiding lattice polygons} or \emph{polygons},
or SAPs, for short.  
These are embeddings of the circle graph in a lattice.  Each vertex has degree 2.
Let $p_n$ be the number of (undirected, unrooted) $n$-edge polygons in ${\mathbb Z}^d$, 
counted modulo translation.  In two dimensions, $p_4 = 1$, $p_6 = 2$, $p_8=7$, etc.  Clearly
$p_{2m+1} = 0$.  Let $p_n(v,s)$ be the number of $n$-edge polygons with all
vertices having $x_d \ge 0$, $v+2$ visits ($v \ge 0$)
and span in the $x_d$-direction equal to $s$, counted modulo translation parallel to the surface.

Define the partition function 
\begin{equation}
P_n(a,y) = \sum_{v \ge 0,s \ge 0} p_n(v,s) a^vy^s. 
\label{eqn:polygonpf}
\end{equation}
The fugacities are again interpreted as in equation (\ref{eqn:fugacities}) and now
the force is applied at the highest vertex.

In this paper we show that in $d\geq 3$ dimensions this model has 
a thermodynamic limit with free energy defined by
\begin{equation}
\psi_0(a,y) = \lim_{n\to\infty} \sfrac{1}{n} \log P_n(a,y) .
\end{equation}
In addition, we show that, in this case, for $a\geq a_c^0$ (where $a_c^0$
is the critical adsorption fugacity for adsorbing polygons) and $y\geq 1$,
\begin{equation}
\psi_0(a,y) = \max [ \kappa_0(a),\lambda_0(y) ]
\end{equation}
where $\kappa_0(a)$ is the free energy of adsorbing polygons in the absence of a
pulling force, and $\lambda_0(y)$ is the free energy of a pulled polygon in the
absence of an adsorption fugacity.  The resulting phase diagram of this model is
sketched in Figure \ref{figure22}.   We know that the phase boundary between the  
free phase and the adsorbed phase is a vertical line, and the phase boundary between the
free phase and the ballistic phase is a horizontal line.

In $d=2$ dimensions, our results are less complete.

\section{Polygons adsorbing at a surface}
\label{sec:pureadsorb}

If we consider polygons adsorbing at a surface with no applied force  ($y=1$) we have the 
following theorem:
\begin{theo}[Soteros]
The thermodynamic limit 
\begin{equation} 
\lim_{n\to\infty} n^{-1} \log P_n(a,1) \equiv \kappa_0(a) < \infty
\end{equation}
exists.  Moreover, $\kappa_0(a)$ is a convex function of $\log a$ and hence 
is continuous and differentiable almost everywhere.
\label{theo:Soteros1}
\end{theo}
The proof of this theorem follows from results in \cite{Soteros}, in Sections 3 and 4.

When $d \ge 3$ it is relatively straightforward to show that $\kappa_0(a) = \kappa (a)$ so the location
of the adsorption transition is the same as that of walks.   See
Section 3 of  \cite{Soteros} for a proof when $d=3$ that can be extended to $d > 3$
without much difficulty.
When $d=2$ the situation is slightly different because a polygon cannot lie 
entirely in the confining line.  We have the following theorem.
\begin{theo}[Soteros]
Suppose that $d=2$.  When $a \le 1$ then $\kappa_0(a) = \kappa_0(1)
=\kappa_2$.  When $a > 1$
\begin{equation}
 \max[\kappa_2, (1/2) \log a] \le \kappa_0(a) \le \kappa_2 + (1/2) \log a.
\end{equation}
\label{theo:Soteros2}
\end{theo}
A proof is given in Section 4 of \cite{Soteros}.

This theorem implies that there is an adsorption transition at $a=a_c^0$ where 
\begin{equation}
1 \le a_c^0 \le \exp[2 \kappa_2]
\label{eqn:inequalities}
\end{equation}
and, with a little more effort, the upper bound 
can be made strict.  By deleting a suitable edge each polygon can be converted 
into a terminally attached self-avoiding walk so $\kappa_0(a) \le \kappa (a)$ 
and this implies that $a_c^0 \ge a_c$.  Since we know 
\cite{HTW1982} that $a_c > 1$ this implies that $a_c^0  > 1$ so both
inequalities in  (\ref{eqn:inequalities}) are strict.  It is an open question as to whether the two 
critical points are identical or not.

\section{Polygons pulled from a non-interacting surface}
\label{sec:purepull}

In this section we consider a polygon attached to an impenetrable surface and pulled 
away from the surface.  There is no attractive interaction so $a=1$.  The force is conjugate 
to the span ($s$) of the polygon in the $x_d$-direction and the partition function is 
\begin{equation}
P_n(1,y) = \sum_{v\ge 0,s \ge 0} p_n(v,s) y^s.
\end{equation}
The existence of the limit 
\begin{equation}
\lim_{n\to\infty} n^{-1} \log P_n(1,y) = \lambda_0(y)
\end{equation}
is established in \cite{Rensburg2008}.  In addition we know \cite{Rensburg2008}
that $\lambda_0(y)$ is 
a convex function of $\log y$ (and hence continuous) and, for $y \ge 1$, 
$\lambda_0(y)$ satisfies the bounds
\begin{equation}
\max [\lambda_0(1), (1/2) \log y ] \le \lambda_0(y) \le \lambda_0 (1) + (1/2) \log y.
\label{eqn:lambdabound}
\end{equation}
Note that $\lambda_0(1) = \kappa_d$.

\subsection{Polygons in a slit or slab}
\label{ssectionA}

It will be useful to consider polygons with at least one vertex  in the bottom boundary
of a slit or slab of width $w$ in the hypercubic lattice.  Let $\pi_n(w)$ be the number of
polygons of length $n$ with highest vertices at height $w$, so
that they fit in a slab or slit of width $w$ (see Figure \ref{fig1}), counted modulo translation parallel to the 
confining boundaries.

The generating function of this model is
\begin{equation}
\Pi(t,y) = \sum_{n=0}^\infty \sum_{w=0}^{n/2} \pi_n(w)\,y^wt^n ,
\end{equation}
where $\pi_n(w) = \sum_v p_n(v,w)$.
If $0 \leq y \leq 1$, by monotonicity, 
\begin{equation}
G(t) = \Pi(t,1) \geq \Pi(t,y) ,
\end{equation}
where $G(t)$ is the generating function of polygons in a half-lattice.
Then it is known that $G(t)$ is singular at $t=t_c = \frac{1}{\mu} = e^{-\kappa_d}$.

Suppose that $d \ge 3$.  Then it is known 
\begin{equation}
\lim_{n\to \infty} n^{-1} \log \pi_n(w) = \log \mu_w
\end{equation}
exists where $\mu_w$ is the growth constant for self-avoiding walks in a 
slab of width $w$ \cite{HamWhitt1985}.  It is proved there that 
$\mu_w < \mu_{w+1}$ and  $\lim_{w \to \infty} \mu_w = \mu$.
$\Pi(t,y)$ is bounded from below by any term in its defining series:
\begin{equation}
\Pi (t,y) \geq y^w\sum_{n=2w}^\infty \pi_n(w)\,t^n  = y^w\, G_w(t).
\end{equation}
$G_w(t)$ is singular at $t_w = \frac{1}{\mu_w}$.

The above gives
\begin{equation}
G(t) \geq \Pi(t,y) \geq y^w\, G_w(t),\quad \hbox{if $y \leq 1$, for every $w> 0$}.
\end{equation}
The generating functions $G(t)$ and $G_w(t)$ have radius of convergence
$\frac{1}{\mu}$ and $\frac{1}{\mu_w}$ respectively, where $\mu_w < \mu$.
This shows, that for any fixed $0<y\leq1$, the radius of convergence of
$\Pi (t,y)$ is $\frac{1}{\mu_0(y)}$, where
\begin{equation}
\mu \geq \mu_0(y) \geq \mu_w .
\end{equation}
Since $\mu_w \nearrow \mu$ as $w\to\infty$, this shows that
\begin{equation}
\mu_0(y) = \mu , \quad\hbox{for all $0<y\leq 1$}
\end{equation}
for all $d \ge 3$.

It remains to consider the case of $d=2$.  There we know that the growth constant 
$\mu_w^0$ for polygons in a slab of width $w$ is strictly less than that of walks in a 
slab of width $w$, \emph{i.e.} $\mu_w^0 < \mu_w$ 
for all $w < \infty$ \cite{Soteros1988}.  The same argument shows 
that
\begin{equation}
\mu \ge \mu_0(y) \ge \mu_w^0
\label{eqn:2dbounds}
\end{equation}
and we give some  properties of $\mu_w^0$ in the next theorem.

\begin{theo}
For polygons in a two dimensional slit $\mu_w^0$ is an increasing function of 
$w$ and  $\sup_w \mu_w^0 = \mu$.
\end{theo}
\Pr
Every polygon with $n$ edges and span $w$ can be converted to a polygon with $n+2$ 
edges and span $w+1$ so $\pi_{n+2}(w+1) \ge \pi_n(w)$, from which we have 
$\mu_{w+1}^0 \ge \mu_w^0$.   The existence of the limit $\lim_{w \to \infty} \mu_w^0$
and the fact that it is equal to $\sup_w \mu_w^0$ and that this in turn is equal to $\mu$
is a consequence of the arguments in Sections 4 and 6 of \cite{HamWhitt1985}.
\qed

Since $\sup_w \mu_w^0 = \mu$ (by the above Theorem) it follows from 
(\ref{eqn:2dbounds}) that $\mu_0(y) = \mu$ for all $y \le 1$ in two dimensions, and 
therefore for all $d \ge 2$.

\vspace{0.1in}

\subsection{Polygons pulled at a middle vertex}

Consider $n$-edge polygons ($n$ is automatically even)
with at least one vertex in the adsorbing plane and with all vertices in or on one 
side of this plane.  Translate the polygon so that the lexicographically first vertex
in the surface is at the origin.   The \emph{middle vertex} 
is the vertex joined to the origin by two sub-walks each of length $n/2$.

Let $c_n^+(h)$ be the number of positive walks from the origin with endpoint at height $h$ above the adsorbing
plane, and $p_n^+(h)$ be the number of polygons with middle vertex
at height $h$.  Note that  $\sum_h c_n^+(h) = c_n^+$.
Let $C_n^+(y)$ be the partition function of $c_n^+(h)$ and $P_n^+(y)$ be the partition function of $p_n^+(h)$,
with growth constants $\mu_c^+(y)$ and $\mu_p^+(y)$.   The free energies are $\lambda^+(y) = \log \mu_c^+(y)$ and $\lambda_p^+(y)
= \log \mu_p^+(y)$.  The partition function $P_n^+(y)$ is the partition function of 
polygons pulled at their middle vertex and $C_n^+(y)$ is the corresponding partition function 
for walks pulled at their last vertex.

We define the generating functions
\begin{equation}
W^+(t,y) = \sum_n C_n^+(y) t^n \quad \quad P^+(t,y) = \sum_n P_n^+(y) t^n
\end{equation}
and we write their radii of convergence as $t_c^+(y)$ and $t_p^+(y)$.  

We now define similar quantities for bridges.  A bridge is a positive walk that takes its first step 
away from the adsorbing plane, never returns to the adsorbing plane and whose last vertex is
in the top plane of the walk.  Let $b_n(h)$ be the number of $n$-edge bridges of height $h$.  The partition 
function and generating function are defined as 
\begin{equation}
B_n(y) = \sum_h b_n(h) y^h , \quad \quad B(t,y) = \sum_n B_n(y) t^n.
\end{equation}

A bridge can be doubly unfolded in the first coordinate direction so that 
the origin is left-most and the first coordinate of the last vertex is 
right-most \cite{HammersleyWelsh}.  See Figure \ref{figure2}.

Let $b_n^\dagger(h)$ be the number of doubly unfolded bridges with $n$ edges and height $h$.
Then
\[ b_n^\dagger(h) \leq b_n (h) \leq e^{o(n)} b_n^\dagger(h). \]
Their partition functions are related by the inequalities, 
\begin{equation}
B_n^\dagger(y) \leq B_n(y) \leq e^{o(n)} B_n^\dagger(y), 
\end{equation}
and so
$\lim_{n\to\infty} (B_n^\dagger(y))^{1/n} = \mu_c^B(y)$ where $\mu_c^B(y) = 1/t_c^B(y)=1/t_c^+(y)$.   This last 
equality follows from Lemma 1  and Theorem 3 in \cite{RensburgWhittington2016a}.


\subsection{A connection between polygon and walk partition functions}


Clearly
$t_c^+(y) \leq t_p^+(y)$ for all $y>0$ and $t_c^+(1) = t_p^+(1) = \mu^{-1}$.
If $y>1$ then $P_n^+ (y) \leq p_n y^{n/2}$ and $C_n^+(y) \geq y^n$.  Taking powers $1/n$ and then letting $n\to\infty$
gives $\mu_p^+(y) \leq \mu\,\sqrt{y} < y \leq \mu_c^+(y)$ for $y > \mu^2$, or
\begin{equation}
 t_p^+(y) > t_c^+(y), \quad\hbox{for $y > \mu^2$}. 
 \label{eqn:polywalk1inequality}
 \end{equation}
We next show how this inequality can be strengthened.

\begin{theo}
$$t_p^+(y^2) \ge t_c^+(y), \: \mbox{for all } y \ge 1.$$
\label{theo:polywalk}
\end{theo}
\Pr
By cutting the polygon at its middle vertex (where the force is applied) into two walks, $p_n^+ (h) \leq (c_{n/2}^+(h))^2$, so that
\begin{equation}
\fl  \quad
 P_n^+(y) = \sum_h p_n^+(h) y^h \leq \sum_h \L c_{n/2}^+(h) y^{h/2} \R^2
\leq \L \sum_h c_{n/2}^+(h) y^{h/2} \R^2 = (C_{n/2}^+(\sqrt{y}))^2 . 
\end{equation}
Take the power $1/n$ and let $n\to\infty$ to obtain $\mu_p^+(y) \leq \mu_c^+(\sqrt{y})$.  That is
\[ t_p^+(y^2) \geq t_c^+(y), \quad\hbox{for $y \geq 1$}.  \]
\qed

A corollary of this Theorem is as follows:
The free energy of pulled walks is $-\log t_c^+(y)$, and this is strictly increasing with $y>1$, since the model is ballistic.
That is, $t_c^+(y) > t_c^+(y^2)$ if $y>1$.  This shows with the above that
$t_p^+(y^2) \geq t_c^+(y) > t_c^+(y^2)$ for $y>1$, or $t_p^+(y) > t_c^+(y)$ whenever $y>1$.  This strengthens (\ref{eqn:polywalk1inequality}).

\def\Vec#1{\overset{\to}{#1}}

We next prove the corresponding inequality in the other direction.  We obtain a lower bound by constructing polygons 
by concatenating four doubly unfolded bridges.  
The unfolded bridge has a width $\Vec{w}$ and a height $h$, where $\Vec{w}$ is the vector of 
widths in all except the vertical direction, and may look as illustrated in Figure
\ref{figure2}.

\begin{figure}[t]
\beginpicture
\setcoordinatesystem units <1.5pt,1.5pt>
\setplotarea x from -130 to 90, y from -10 to 80
\setplotarea x from -50 to 50, y from 0 to 80

\arrow <5pt> [.2,.67] from 60 0 to 60 80
\put {\large $h$} at 65 40

\arrow <5pt> [.2,.67] from -50 -10 to 50 -10
\put {\large $\overset{\longrightarrow}{w}$} at 0 -15 

\color{Tan}
\grid 10 8 
\color{Tan}
\setplotsymbol ({$\bullet$})
\plot -50 -3 50 -3 /

\color{Blue}
\plot -50 0 -50 10 -40 10 -40 20 -50 20 -50 30 -40 30 -30 30 -30 20 -30 10 -20 10 -20 20 
-20 30 -20 40 -20 50 -20 60 -30 60 -30 70  -30 80 -20 80 -20 70 -10 70 0 70 10 70 10 60 10 50 
10 40 0 40 0 30 10 30 20 30 20 40 20 50 30 50 40 50 50 50 50 60 40 60 40 70 50 70 
50 80 /
\color{NavyBlue}
\multiput {\LARGE$\bullet$} at 
-50 0 -50 10 -40 10 -40 20 -50 20 -50 30 -40 30 -30 30 -30 20 -30 10 -20 10 -20 20 
-20 30 -20 40 -20 50 -20 60 -30 60 -30 70 -30 80 -20 80 -20 70 -10 70 0 70 10 70 10 60 10 50 
10 40 0 40 0 30 10 30 20 30 20 40 20 50 30 50 40 50 50 50 50 60 40 60 40 70 50 70 
50 80 /

\color{red}
\multiput {\huge$\bullet$} at -50 0 50 80 /

\color{black} \normalcolor

\endpicture
\caption{A doubly unfolded bridge.  The vertex in the surface is in the left-most plane and the last vertex
is in the right-most plane.  The bridge steps away from the surface at its first step and 
never returns.}
\label{figure2}
\end{figure}

%
%

Next, we have to choose values for $h$ and $\Vec{w}$.  Let $b_n^{\dagger}(h,\Vec{w})$ be the number of  $n$-edge 
doubly unfolded bridges with width $\Vec{w}$ and height $h$.  
In the partition function for doubly unfolded bridges, $B_n^\dagger (y) = \sum_{h,\Vec{w}} b_n^\dagger (h,\Vec{w}) y^h$
there are \textit{most popular values} of $h$ and $\Vec{w}$ (these are functions of $y$); denote them by $h^*$ and $\Vec{w}^{\,*}$.
Then $b_n^\dagger(h^*,\Vec{w}^{\,*}) y^{h^*}$ is a largest term in $B_n^\dagger(y) = \sum_{h,\Vec{w}} b_n^\dagger(h,\Vec{w}) y^h$, so that
\begin{equation}
 b_n^\dagger(h^*,\Vec{w}^{\,*}) y^{h^*} \leq B_n^\dagger(y) \leq (n+1)^d\, b_n^\dagger(h^*,\Vec{w}^{\,*}) y^{h^*}.
 \label{eqn:unfoldedbridge}
 \end{equation}
Taking powers $1/n$, letting $n\to\infty$, and noting that $(B_n^\dagger(y))^{1/n} \to \mu_c^B(y)$, it follows that
\begin{equation} \lim_{n\to\infty} (b_n^\dagger(h^*,\Vec{w}^{\,*})y^{h^*})^{1/n} = \mu^B_c(y) = 1/t_c^B(y)= 1/t_c^+(y). 
\label{eqn:bridgemu}
\end{equation}

\begin{figure}[t]
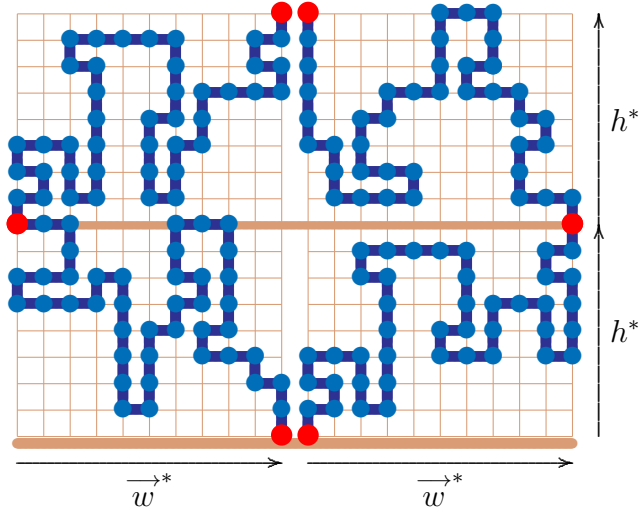

\beginpicture

\setcoordinatesystem units <1.0pt,1.0pt>
\setplotarea x from -130 to 170, y from -10 to 160

\setplotarea x from -50 to 50, y from 0 to 160

\arrow <5pt> [.2,.67] from -50 -10 to 50 -10
\put {\large $\overset{\longrightarrow}{w}^*$} at 0 -20
\arrow <5pt> [.2,.67] from 60 -10 to 160 -10
\put {\large $\overset{\longrightarrow}{w}^*$} at 110 -20

\arrow <5pt> [.2,.67] from 170 0 to 170 80
\put {\large $h^*$} at 180 40
\arrow <5pt> [.2,.67] from 170 80 to 170 160
\put {\large $h^*$} at 180 120

\color{Tan}
\grid 10 16 
\color{Tan}
\setplotsymbol ({\scriptsize$\bullet$})
\plot -50 80 160 80 /
\setplotsymbol ({$\bullet$})
\plot -50 -3 160 -3 /

\color{Blue}
\plot -50 80 -40 80 -30 80 -30 70 -30 60 -40 60 -50 60 -50 50 -40 50 -30 50 -20 50 
-20 60 -10 60 -10 50 -10 40 -10 30 -10 20 -10 10  0 10 0 20 0 30 0 40 10 40 10 50 
20 50 20 60 10 60 10 70 10 80 20 80 30 80 30 70 30 60 30 50 30 40 20 40 20 30  
30 30 40 30 40 20 50 20 50 10 50 0 /
\color{NavyBlue}
\multiput {\LARGE$\bullet$} at 
-50 80 -40 80 -30 80 -30 70 -30 60 -40 60 -50 60 -50 50 -40 50 -30 50 -20 50 
-20 60 -10 60 -10 50 -10 40 -10 30 -10 20 -10 10  0 10 0 20 0 30 0 40 10 40 10 50 
20 50 20 60 10 60 10 70 10 80 20 80 30 80 30 70 30 60 30 50 30 40 20 40 20 30  
30 30 40 30 40 20 50 20 50 10 50 0 /

\color{red}
\multiput {\huge$\bullet$} at -50 80 50 0 /

\color{black} \normalcolor

\setcoordinatesystem units <1.0pt,1.0pt> point at -110 0

\setplotarea x from -50 to 50, y from 0 to 160

\color{Tan}
\grid 10 16

\color{Blue}
\plot -50 0 -50 10 -40 10 -40 20 -50 20 -50 30 -40 30 -30 30 -30 20 -30 10 -20 10 
-20 20 -20 30 -20 40 -20 50 -20 60 -30 60 -30 70 -20 70 -10 70 0 70 10 70 10 
60 10 50 10 40 0 40 0 30 10 30 20 30 20 40 20 50 30 50 40 50 40 40 40 
30 50 30 50 40 50 50 50 60 40 60 40 70 50 70 50 80 /
\color{NavyBlue}
\multiput {\LARGE$\bullet$} at 
-50 0 -50 10 -40 10 -40 20 -50 20 -50 30 -40 30 -30 30 -30 20 -30 10 -20 10 
-20 20 -20 30 -20 40 -20 50 -20 60 -30 60 -30 70 -20 70 -10 70 0 70 10 70 10 
60 10 50 10 40 0 40 0 30 10 30 20 30 20 40 20 50 30 50 40 50 40 40 40 
30 50 30 50 40 50 50 50 60 40 60 40 70 50 70 50 80  /

\color{red}
\multiput {\huge$\bullet$} at -50 0 50 80 /

\color{black} \normalcolor

\setcoordinatesystem units <1.0pt,1.0pt> point at 0 -80

\setplotarea x from -50 to 50, y from 0 to 80

\color{Blue}
\plot -50 0 -50 10 -40 10 -40 20 -50 20 -50 30 -40 30 -30 30 -30 20 -30 10 -20 10 -20 20 
-20 30 -20 40 -20 50 -20 60 -30 60 -30 70 -20 70 -10 70 0 70 10 70 10 60 10 50 
10 40 0 40 0 30 0 20 0 10 10 10 10 20 10 30 20 30 20 40 20 50 30 50 40 50 50 50 
50 60 40 60 40 70 50 70 50 80 /
\color{NavyBlue}
\multiput {\LARGE$\bullet$} at 
-50 0 -50 10 -40 10 -40 20 -50 20 -50 30 -40 30 -30 30 -30 20 -30 10 -20 10 -20 20 
-20 30 -20 40 -20 50 -20 60 -30 60 -30 70 -20 70 -10 70 0 70 10 70 10 60 10 50 
10 40 0 40 0 30 0 20 0 10 10 10 10 20 10 30 20 30 20 40 20 50 30 50 40 50 50 50 
50 60 40 60 40 70 50 70 50 80  /

\color{red}
\multiput {\huge$\bullet$} at -50 0 50 80 /

\color{black} \normalcolor

\setcoordinatesystem units <1.0pt,1.0pt> point at -110 -80

\setplotarea x from -50 to 50, y from 0 to 80

\color{Blue}
\plot -50 80 -50 70 -50 60 -50 50 -50 40 -50 30 -40 30 -40 20 -40 20 -40 10 
-30 10 -20 10 -10 10 -10 20 -20 20 -30 20 -30 30 -30 40 -20 40 -20 50 -10 50 
0 50 0 60 0 70 0 80 10 80 20 80 20 70 20 60 20 60 10 60 10 50 20 50 
30 50 30 40 40 40  40 30 30 30 30 20 30 10 40 10 50 10 50 0 / 
\color{NavyBlue}
\multiput {\LARGE$\bullet$} at 
-50 80 -50 70 -50 60 -50 50 -50 40 -50 30 -40 30 -40 20 -40 20 -40 10 
-30 10 -20 10 -10 10 -10 20 -20 20 -30 20 -30 30 -30 40 -20 40 -20 50 -10 50 
0 50 0 60 0 70 0 80 10 80 20 80 20 70 20 60 20 60 10 60 10 50 20 50 
30 50 30 40 40 40  40 30 30 30 30 20 30 10 40 10 50 10 50 0  /

\color{red}
\multiput {\huge$\bullet$} at -50 80 50 0 /

\color{black} \normalcolor

\endpicture
\caption{Four doubly unfolded bridges can be concatenated to form a polygon.  The bridges are selected 
so that their heights and widths have their most popular values.}
\label{figure3}
\end{figure}

\begin{theo}
$$t_p^+(y^2) = t_c^+(y) \quad \hbox{for all $y \ge 1$}.$$
\end{theo}
\Pr
Because of Theorem \ref{theo:polywalk} above we only need to prove an inequality in one direction.
By reflecting and rotating unfolded bridges of most popular widths and heights
they can be concatenated to form a  polygon as in Figure \ref{figure3}.

%
%
%
%
%
%
%
%

This arrangement gives a lower bound on polygons of height $2h^*$ and of length $4n+2$ which gives a lower bound
on the polygon partition function:
\begin{equation}
 \L b_n^\dagger(h^*,\Vec{w}^{\,*}) y^{h^*} \R^{\!4} \leq   p^+_{4n+2}(2h^*) y^{4h^*}   \leq P^+_{4n+2}(y^2) .
 \label{eqn:bridgepol}
 \end{equation}
Take the power $1/4n$ and then let $n\to\infty$.  The left hand side goes to $1/t_c^+(y)$, and the right hand side goes to 
$\mu_p^+(y^2) = 1/t_p^+(y^2)$.  This shows that
\[ t_p^+(y^2) \leq t_c^+(y). \]
Since we already know that $t_p^+(y^2) \geq t_c^+(y)$ the result is that
\begin{equation}
 t_p^+(y^2) = t_c^+(y) .
\label{eqn1}  
\end{equation}
This shows that polygons also become ballistic at $y_c=1$.
\qed

We turn our attention now to the case where $y \le 1$.  The construction in Figure \ref{figure3},
together with equations (\ref{eqn:unfoldedbridge}) and (\ref{eqn:bridgepol}), shows that
\begin{eqnarray}
\frac{1}{(n+1)^{4d}}(B_n^{\dagger}(y))^4 & \le & (b_n^{\dagger}(h^*,\Vec{w}^*)y^{h^*})^4 \nonumber \\
& \le & p_{4n+2}^+(2h^*)y^{4h^*} \le P_{4n+2}^+(y^2).
\end{eqnarray}
Recall that $h^*$ and $\Vec{w}^*$ are the most popular values of the height and width.
We know that, for $y \le 1$, $t_c^+(y) = t_c^B(y)= 1/\mu$ \cite{RensburgWhittington2016a}.
Hence $1/\mu \ge t_p^+(y^2) \ge t_c^+(y^2) = 1/\mu$ for all $y \le 1$.  This proves the following
theorem:
\begin{theo}
When $y \le 1$, $ t_p^+(y) = \frac{1}{\mu}$.
\label{theo:QQ}
\end{theo}

\section{Polygons pulled from their top plane}
\label{section5}



Define $p_n(\ell)$ to be the number of polygons, with at least one vertex in $x_d=0$ 
and with \textit{highest vertices} at height $\ell$.  A model of pulled polygons,
where the highest vertices are pulled vertically by a force $f$, is defined by
the partition function
\begin{equation}
P_n (y) = \sum_{\ell\geq 0} p_n(\ell)\, y^\ell.
\end{equation}
Here, the activity $y=e^{f /k_BT}$ is introduced and is equal to the exponential of the
reduced pulling force.  The generating function of this model is given by
$P(t,y)$ and its radius of convergence is denoted $t_p(y)$.

It remains to relate the limiting free energy of this model to that of polygons
pulled in the middle, considered above.  Note that the midpoint
of a polygon of height $\ell$ is itself at most at height $\ell$.  Thus, assuming
that $y\geq 1$, $P_n^+(y) \leq P_n(y)$.  This shows that
\begin{equation}
\lim_{n\to\infty} \frac{1}{n} \log P_n^+(y) \leq \liminf_{n\to\infty}
\frac{1}{n} \log P_n(y) .
\label{eqn2}  
\end{equation}
This, in particular, shows that $t_c^+(\sqrt{y}) = t_p^+(y) \geq t_p(y)$.

Existence of the free energy will now be shown by bounding the limiting supremum.

\begin{figure}[t]
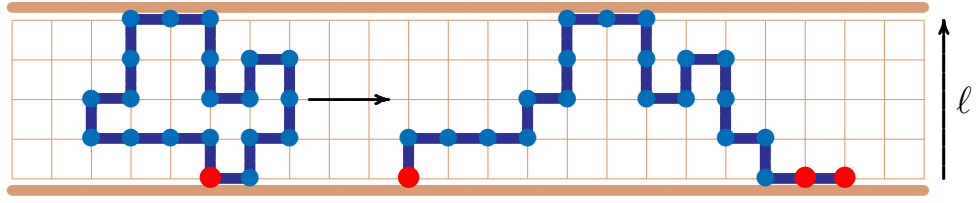

\beginpicture
\setcoordinatesystem units <1.5pt,1.5pt>
\setplotarea x from -150 to 130, y from -10 to 50
\setplotarea x from -100 to 130, y from 0 to 40

\color{Tan}
\grid 23 4 
\color{Tan}
\setplotsymbol ({$\bullet$})
\plot -100 -3 130 -3 /
\plot -100 43 130 43 /

\color{Blue}
\plot -50 0 -50 10 -60 10  -70 10 -80 10 -80 20 -70 20 -70 30 -70 40 
-60 40 -50 40 -50 30  -50 20 -40 20 -40 30 -30 30 -30 20 -30 10 -40 10 -40 0 -50 0  /
\plot 0 0 0 10 10 10 20 10 30 10 30 20 40 20 40 30 40 40 50 40 60 40 
60 30 60 20 70 20 70 30 80 30 80 20 80 10 90 10 90 0 100 0 110 0 /
\color{NavyBlue}
\multiput {\LARGE$\bullet$} at  
-50 0 -50 10 -60 10 -70 10 -80 10 -80 20 -70 20 -70 30 -70 40 
-60 40 -50 40 -50 30  -50 20 -40 20 -40 30 -30 30 -30 20 -30 10 -40 10 -40 0 -50 0
0 0 0 10 10 10  20 10 30 10 30 20 40 20 40 30 40 40 50 40 60 40 
60 30 60 20 70 20 70 30 80 30 80 20 80 10 90 10 90 0 100 0 110 0 /

\color{red}
\multiput {\huge$\bullet$} at -50 0 0 0 100 0 110 0  /

\color{black} \normalcolor
\setplotsymbol ({.})

\arrow <5pt> [.2,.67] from -25 20 to -5 20 
\arrow <5pt> [.2,.67] from 135 0 to 135 40 
\put {\Large$\ell$} at 140 20

\endpicture
\caption{A polygon in a slit or slab unfolded to form a loop.}
\label{figure4}
\end{figure}

Cut the polygon at its lexicographically first vertex in the surface, and unfold it into a 
bridge in the $x_1$-direction  (by
adding a single edge in the horizontal direction).  This is
schematically illustrated in Figure \ref{figure4}.

%

\begin{figure}[t]
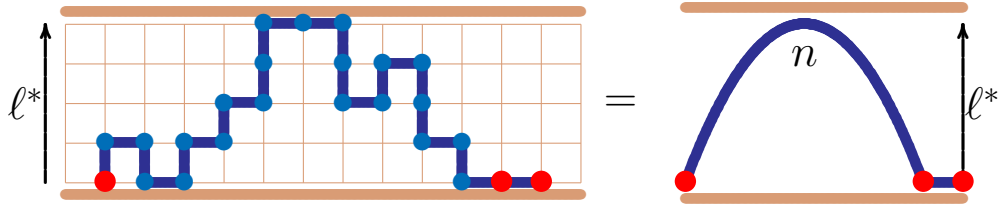

\beginpicture
\setcoordinatesystem units <1.5pt,1.5pt>
\setplotarea x from -50 to 120, y from -10 to 50
\setplotarea x from -10 to 120, y from 0 to 40

\color{Tan}
\grid 13 4 
\color{Tan}
\setplotsymbol ({$\bullet$})
\plot -10 -3 120 -3 /
\plot -10 43 120 43 /

\color{Blue}
\plot 0 0 0 10 10 10 10 0 20 0 20 10 30 10 30 20 40 20 40 30 40 40 50 40 60 40 
60 30 60 20 70 20 70 30 80 30 80 20 80 10 90 10 90 0 100 0 110 0 /
\color{NavyBlue}
\multiput {\LARGE$\bullet$} at  
0 0 0 10 10 10 10 0 20 0 20 10 30 10 30 20 40 20 40 30 40 40 50 40 60 40 
60 30 60 20 70 20 70 30 80 30 80 20 80 10 90 10 90 0 100 0 110 0 /

\color{red}
\multiput {\huge$\bullet$} at 0 0 100 0 110 0  /

\color{black} \normalcolor
\setplotsymbol ({.})

\arrow <5pt> [.2,.67] from -15 0 to -15 40 
\put {\LARGE$\ell^*$} at -20 20

\put {\LARGE$=$} at 130 20 

\put {
\beginpicture
\setplotarea x from -10 to 75, y from -5 to 45

\color{black}
\arrow <5pt> [.2,.67] from 70 0 to 70 40 
\put {\LARGE$\ell^*$} at 75 20
\put {\LARGE$n$} at 30 32

\color{Tan}
\setplotsymbol ({$\bullet$})
\plot 0 -4 70 -4 /  \plot 0 44 70 44 /

\color{Blue}
\plot 60 0 70 0 /
\setquadratic
\plot 0 0 30 40 60 0  /
\setlinear

\color{red}
\multiput {\huge$\bullet$} at 0 0 60 0 70 0 / 

\color{black}

\endpicture } at 180 20

\color{black}
\normalcolor
 
\endpicture
\caption{An unfolded loop with its last step in the surface and with height equal to the most
popular height $\ell^*$.  On the right is a schematic diagram
of this class of loops.}
\label{figure5}
\end{figure}

The polygon is unfolded into a loop of height $\ell$.
Denote the partition function of these (unfolded) loops by $L_n^\ddagger(y)$, 
and it follows that $P_n(y) \leq e^{o(n)} L_n^\ddagger (y)$.  Let the 
number of unfolded loops of 
length $n$ and height $\ell$ be denoted by $l_n^\ddagger(\ell)$.  Then
\begin{equation}
L_n^\ddagger(y) = \sum_{\ell=0}^{n/2} l_n^\ddagger(\ell)\,y^\ell.
\end{equation}
For each value of $y>0$ there is a most popular value of $\ell$ in this summation,
and this is denoted by $\ell^*$ (this is dependent on $y$ and on $n$).  In particular,
\begin{equation}
l_n^\ddagger(\ell^*)\,y^{\ell^*} \leq L_n^\ddagger(y) \leq \sfrac{1}{2}n\, l_n^\ddagger(\ell^*)\,y^{\ell^*}.
\end{equation}
The loops in this most popular class are schematically illustrated 
in Figure \ref{figure5}.

\begin{figure}[t]
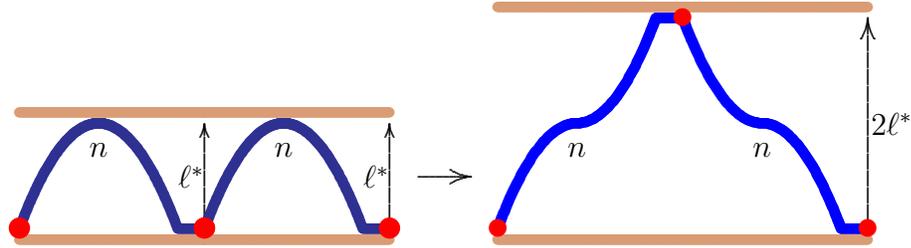

\beginpicture
\setplotarea x from -100 to 205, y from -5 to 45
\setplotarea x from -10 to 205, y from -5 to 45

\arrow <8pt> [.2,.67] from 120 20 to 140 20

\multiput {
\beginpicture
\setplotarea x from -10 to 75, y from -5 to 45

\color{black}
\arrow <5pt> [.2,.67] from 70 0 to 70 40 
\put {\large$\ell^*$} at 65 20
\put {\large$n$} at 30 30

\color{Tan}
\setplotsymbol ({$\bullet$})
\plot 0 -4 70 -4 /  \plot 0 44 70 44 /

\color{Blue}
\plot 60 0 70 0 /
\setquadratic
\plot 0 0 30 40 60 0  /
\setlinear

\color{red}
\multiput {\huge$\bullet$} at 0 0  70 0 / 

\color{black}

\endpicture } at 0 20 70 20 /

\setplotarea x from 150 to 250, y from -5 to 85

\color{black}
\arrow <8pt> [.2,.67] from 290 0 to 290 80 
\put {\large$2\ell^*$} at 299 40
\setsolid
\multiput {\large$n$} at 180 30 250 30  / 

\color{Tan}
\setplotsymbol ({$\bullet$})
\plot 150 -4 290 -4 /  \plot 150 84 290 84 /

\color{blue}
\plot 210 80 220 80 /
\plot 280 0 290 0 /
\setquadratic
\plot 150 0 165 30 180 40 195 50 210 80  /
\plot 220 80 235 50 250 40 265 30 280 0 /
\setlinear

\color{red}
\multiput {\LARGE$\bullet$} at 150 0 220 80 290 0 /

\color{black}
\normalcolor

\endpicture
\caption{Two unfolded loops are concatenated in a slab of height $\ell^*$ and 
then reflected through the top boundary of the slab to obtain a loop in a slab of 
height $2\ell^*$.}
\label{figure6}
\end{figure}

Two loops in this most popular class can be concatenated as illustrated 
schematically in Figure \ref{figure6}.
If the middle part of the concatenated loops is reflected through
the top plane as shown, then a loop of height $2\ell^*$ is obtained
with the property that its middle vertex is also in the top plane.  This 
loop consists of two self-avoiding walks of length $n$ and height $2\ell^*$.
Thus
\begin{eqnarray}
\frac{4}{n^2} L_n^{\ddagger 2}(y) & \leq
\L l_n^\ddagger(\ell^*)\, y^{\ell^*} \R^2 \leq
\L c_n^+(2\ell^*) \, y^{\ell^*} \R^2 \nonumber \\ 
& \leq \sum_\ell \L c_n^+(\ell) \, y^{\ell/2} \R^2 \leq
\L \sum_\ell c_n^+(\ell)\, y^{\ell/2} \R^2 = (C_n^+ (\sqrt{y}))^2 . \label{eqn5.5} 
\end{eqnarray}
That is, since $P_n (y) \leq e^{o(n)} L_n^\ddagger(y)$, 
\begin{equation}
\limsup_{n\to\infty} \frac{1}{n} \log P_n(y) \leq \lim_{n\to\infty} \frac{1}{n} \log C_n^+(\sqrt{y})
\end{equation}
with the result that by equation \Ref{eqn2},
\begin{eqnarray}
- \log t_p^+(y) &= 
\lim_{n\to\infty} \frac{1}{n} \log P_n^+(y) \leq \liminf_{n\to\infty}
\frac{1}{n} \log P_n(y) \nonumber \\
& \leq \limsup_{n\to\infty} \frac{1}{n} \log P_n(y) \leq \lim_{n\to\infty} \frac{1}{n}  \log C_n^+(\sqrt{y})
\nonumber \\
&= - \log t_c^+(\sqrt{y})= - \log t_p^+(y).
\end{eqnarray}
Here, recall that $P_n^+(y)$ is the partition function of polygons pulled in
the middle, $P_n(y)$ is the partition function of polygons pulled in
their highest plane, and $C_n^+(\sqrt{y})$ is the partition function of 
walks pulled at their endpoint, and $y\geq 1$.

By equation \Ref{eqn1} this shows that for polygons pulled in their top plane,
\begin{equation}
\lim_{n\to\infty} \frac{1}{n} \log P_n (y) = - \log t_c^+(\sqrt{y}) = - \log t_p^+(y).
\end{equation}
That is, the free energy of polygons pulled at their middle point is equal to the
free energy of polygons pulled in their top plane if $y\geq 1$.  The case of $y \le 1$ 
follows from Figure \ref{figure3} and the proof of Theorem \ref{theo:QQ}.  This gives the 
following theorem:
\begin{theo}
$$\lambda_0(y) = \lim_{n \to \infty} \frac{1}{n} \log P_n(y) =  - \log t_c^+(\sqrt{y}) = -\log t_p^+(y)$$
for all $ y > 0$.  When $y \le 1$ this is equal to $\log \mu$.  
\end{theo}
For polygons, as well as for walks, there is a phase transition to a ballistic 
phase at $y=1$ but the response of polygons and walks is different in
this ballistic phase.

\section{Polygons pulled from an interacting surface}
\label{sec:ayplane}
In this section we consider the full problem where the polygon interacts with
the surface ($a \ne 1$) and the applied force is pulling the polymer
off the surface in its top plane ($y > 1$).  We derive some results about the 
$a$-dependence of the free energy at fixed $y > 1$ and show that there is a phase transition from 
an adsorbed phase to a ballistic phase at  $a=a_c^0(y)$.

First consider the situation when $y > 0$ and $a \le 1$.  
\begin{theo}
For $y > 0$ and $a \le 1$ the thermodynamic limit
\begin{equation}
\lim_{n\to\infty} n^{-1} \log P_n (a,y) \equiv \psi_0(a,y)
\end{equation}
exists so we have a well defined limiting free energy.  Moreover, in this region
of the $(a,y)$-plane the free energy is independent of $a$ so that $\psi_0(a,y) = \psi_0(1,y) = \lambda_0(y)$.
\end{theo}
\Pr For fixed $y > 0$ and for all $a \le 1$, by monotonicity 
\begin{equation}
P_n (0,y) \le P_n (a,y) \le P_n (1,y).
\label{eqn:Pbounds}
\end{equation}
Consider an $n$-edge polygon $\omega$ with span $s$.  
Suppose that $V(\omega)$ is the vertex of 
$\omega$ in the surface (\emph{i.e.} that is a visit) with lexicographically
first coordinates in the surface.  When $d=2$ there is exactly one edge of $\omega$ 
in the surface that is incident
on $V$.  For $d>2$, if there are two such edges, choose the one with 
lexicographically first mid-point.
Delete this edge in the surface, incident on $V$, translate $\omega$ by unit distance away from the surface,
add two edges to connect the resulting walk to the surface and add an edge in the 
surface to obtain a polygon $\omega'$ with $n+2$ edges and span $s+1$ with exactly two 
visits. 
\begin{equation}
 \sum_v p_n(v,s) \leq  p_{n+2}(0,s+1) 
.
\end{equation}
This implies that 
\begin{equation}
P_n(1,y) \leq \sfrac{1}{y}\, P_{n+2}(0,y) 
\end{equation}
and hence, from equation (\ref{eqn:Pbounds}),
\begin{equation}
y P_{n-2}(1,y) \le P_n (0,y) \le P_n (a,y) \le P_n (1,y).
\end{equation}
Take logarithms, divide by $n$, let $n\to\infty$, 
and the result follows.
 \qed

When $y \ge 1$ and $a \ge 1$ we have two useful lower bounds on the partition function
that follow from monotonicity, namely
\begin{equation}
P_n(a,y) \ge P_n(a,1) \quad \mbox{and} \quad P_n(a,y) \ge P_n(1,y).
\end{equation}
These bounds imply that
\begin{equation}
\liminf_{n\to\infty} n^{-1} \log P_n(a,y) \ge  \max[\kappa_0(a), \lambda_0(y)].
\end{equation}


We can also construct an upper bound similar to the one derived
for the self-avoiding walk model \cite{JvRW2013}.  Let $l_n(h)$ be the number of 
loops (or arches) with $n$ edges and span in the $x_d$-direction equal to $h$.  
Write the partition function as 
\begin{equation}
L_n(y) = \sum_h l_n(h) y^h.
\end{equation}

\begin{theo}
For all $d \ge 2$
\begin{equation}
P_n(a,y) \le y^{-1} \sum_m C_m(a,1)\, L_{n-m+2}(y)
\label{eqn:upperbound}
\end{equation}
\end{theo}
\Pr
Every $n$-edge polygon with span $h$ must have at least one edge with $x_d=h$.
Either the polygon has every edge in $x_d=0$ so that $h=0$, or $h \ge 1$.
If $h=0$ choose the edge that is lexicographically first
and add two edges to the polygon just before and just after the 
distinguished edge in $x_d=0$ to produce a unique edge in $x_d=1$
in a single loop of 3 edges.  If $h \ge 1$
choose the edge in $x_d=h$ that is lexicographically first.  This edge must be 
in a loop with, say, $n-m$ edges.  The remainder of the polygon is a 
walk with $m$ edges.  Add two edges to the loop just before and just after the 
distinguished edge in $x_d=h$ to produce a unique edge in $x_d=h+1$.  This 
unique edge distinguishes the loop of length $n-m+2$ 
from the rest of the polygon.  By noting that the
rest of the polygon is a walk of length $m$, the following inequality is obtained:
\begin{equation}
p_n (v,h) \le \sum_m c_m(v) \, l_{n-m+2}(h+1)
\end{equation}
where $c_m(v) = \sum_h c_m^+(v,h)$.  Now multiply both sides by
$y^h a^v$ and sum over $h$ and $v$ giving (\ref{eqn:upperbound}) which proves
the theorem.
\qed

Define the generating functions
\begin{equation}
\widehat{L}(t,y) = \sum_n L_n(y)t^n \quad \mbox{and} \quad \widehat{C}(t,a) = \sum_n C_n^+(a,1)t^n.
\end{equation}
By the convolution theorem 
\begin{equation}
\sum_n \sum_m C_m(a,1) L_{n-m+2}(y) t^n \le t^{-2}  \widehat{C}(a,t)  \widehat{L}(y,t) = t^{-2} 
\widehat{B}(a,y,t).
\end{equation}
The radius of convergence of $\widehat{C}(t,a)$ is  $t_1(a)= \exp[-\kappa (a)]$ and the 
radius of convergence of $\widehat{L}(t,y)$ is  $t_2(y)$ so the radius of 
convergence of $\widehat{B}(a,y,z)$ is  $\min[t_1(a),t_2(y)]$.  This implies that
\begin{equation}
\limsup_{n\to\infty} n^{-1} \log P_n(a,y) 
\le \max[\kappa (a), \liminf_{n\to\infty} n^{-1} \log L_n(y)]
\label{eqn6.13}
\end{equation}
or, roughly, the free energy of pulled polygons interacting with the 
surface is bounded above by the maximum of the free energy of 
walks interacting with the surface and pulled loops.

By the results in Section \ref{section5} (see equation \Ref{eqn5.5}) it follows 
\begin{equation}
e^{o(n)}P_n(y) \leq L_n^\ddagger (y) \leq e^{o(n)}C_n^+(\sqrt{y}) = e^{o(n)} P_n (y) .
\end{equation}
Taking logarithms, dividing by $n$, and letting $n\to\infty$, it follows that
\[ \lim_{n\to\infty} \sfrac{1}{n} \log L_n^\ddagger (y) = -\log t_p(y) . \]
But, by unfolding loops, it follows that $e^{o(n)} L_n(y) \leq L_n^\ddagger(y) \leq
L_n(y)$.  Thus
\begin{equation}
\lim_{n\to\infty} \sfrac{1}{n} \log L_n(y) = -\log t_p(y) .
\end{equation}
In equation \Ref{eqn6.13} this gives
\begin{equation}
\limsup_{n\to\infty} n^{-1} \log P_n(a,y) 
\le \max[\kappa (a), \lambda_0(y) ]
\label{eqn6.16}
\end{equation}

When $d > 2$ we know that  $\kappa_0(a) = \kappa (a)$ 
\cite{Soteros} so the above result can be replaced with 
\begin{equation}
\lim_{n\to\infty} n^{-1} \log P_n(a,y)  = \max[\kappa_0 (a), \lambda_0(y)], \quad d \ge 3.
\label{eqn:phaseboundarycondition}
\end{equation}
This gives a complete characterization of the phase boundary when $d\geq 3$ and we state this as a theorem.
\begin{theo}
When $d \ge 3$ the phase boundary between the ballistic and adsorbed phases 
for $y \ge 1$ is determined by the solution of the equation $\kappa_0(a) = \lambda_0(y)$.
\end{theo}
\Pr
This follows immediately from (\ref{eqn:phaseboundarycondition}).
\qed

Since $\kappa_0(a)=\kappa (a)$ \cite{Soteros} and $\lambda_0(y) = \lambda(\sqrt{y})$ the
phase boundary is determined by the properties of the self-avoiding
walk problem.  We know the asymptotics of both $\kappa (a)$ \cite{Rychlewski}
and $\lambda(y)$ \cite{JvRW2013} so we know the behaviour of the phase boundary for polygons at large
values of $a$.  We can switch into the force-temperature plane and this corresponds to the 
behaviour at small values of the temperature.  In particular, the critical force - temperature curve 
is reentrant for all $d > 2$.  The phase transition between the ballistic and adsorbed phases is 
first order, except perhaps at $(a_c^0,1)$.  This follows \emph{mutatis mutandis} from the 
arguments in \cite{Guttmann2014} for the self-avoiding walk model.

When $d=2$ we know that 
\begin{eqnarray}
\max[\kappa_0 (a), \lambda_0(y)]
&\leq \liminf_{n\to\infty} n^{-1} \log P_n(a,y) \nonumber \\
&\leq \limsup_{n\to\infty} n^{-1} \log P_n(a,y)
\leq \max[\kappa (a), \lambda_0(y)] .
\end{eqnarray}

Unlike the self-avoiding walk problem \cite{JvRW2013} we do not have a 
precise condition for locating the phase boundary when $d=2$, since we only 
have lower and upper bounds on the free energy when $a>1$ and $y>1$.  

\section{Pushing a polygon towards an interacting surface}
\label{sec:pushing}

We now consider the situation when $d\geq 3$ and when $0\leq a\leq a_c^0$ and $0<y \leq 1$.  First consider
$0 \leq a \leq 1$ and $0 < y \leq 1$.  By theorem 8 we know that 
\begin{equation}
\psi_0(a,y) = \psi_0(1,y) .
\end{equation}
By theorem 7 it follows that 
$\lambda_0(y) = \psi_0(1,y) = \psi(1,\sqrt{y}) = \lambda(\sqrt{y})$ (see the discussion of the walk 
problem in the introduction).  But $\lambda(\sqrt{y}) = \log \mu$ for $0 < y \leq 1$
\cite{RensburgWhittington2016a}.  Thus $\psi_0(a,y) = \log \mu$ for $0\leq a\leq 1$ and $0<y \leq 1$.

Since $\psi_0(a,y)$ is convex in each of its variables, the critical curve $y_c(a)$ in Figure
\ref{figure22} is a non-decreasing function of $a$.  We now show why $y_c(a)$ has a jump
discontinuity at $a=a_c^0$ so that the phase boundary between the free and adsorbed phases
is a vertical line segment in Figure \ref{figure22}.  Take $a=a_c^0-\epsilon$ for any
$\epsilon>0$.  At $y=1$, $\psi_0(a,1) = \log \mu$.  But $\psi_0(a,y)$ is monotone 
non-decreasing in $y$ so it cannot be greater than $\log \mu$ for $y<1$, so
that $\psi_0(a,y) = \log \mu$ since it cannot be smaller than $\log \mu$.

We now give an alternative proof that the phase boundary between the free and adsorbed phases
is a vertical line segment in Figure \ref{figure22}.   In fact this argument proves considerably 
more and essentially completes our knowledge of the phase diagram
when $d \ge 3$.  We shall need some preliminary lemmas.  Let $\pi_n^w(a) = \sum_v p_n(v,w)a^v$.
By a concatenation argument we can show that the limit
\begin{equation}
\kappa_0^w(a) = \lim_{n\to\infty} n^{-1} \log \pi_n^w(a)
\end{equation}
exists.  For instance, one can use a modification of the concatenation construction used in Section
4 of \cite{HamWhitt1985} coupled with a generalized supermultiplicative inequality.  Let
$\widehat{\pi}_n^w(a) = \sum_{u \le w}\pi_n^u(a)$.  This sum includes all polygons with span at most $w$.

\begin{lemm}
When $d \ge 2$, 
$\kappa_0^w(a) \le \kappa_0^{w+1}(a)$.
\end{lemm}
\Pr 
Each polygon contributing to the sum $\pi_n^w(a) = \sum_v p_n(v,w)a^v$ has at least one 
edge in the hyperplane $x_d=w$.  If there is more than one such edge, take the one with 
lexicographically first midpoint.  Translate this edge unit distance into the hyperplane
$x_d=w+1$ and add two edges to reconnect the polygon.  Then 
\begin{equation}
\pi_n^w(a) \le \pi_{n+2}^{w+1}(a)
\end{equation}
and taking logarithms, dividing by $n$ and letting $n \to \infty$ completes the proof.
\qed

\begin{lemm}
The exponential growth rate of $\widehat{\pi}_n^w(a)$ is identical to that 
of $\pi_n^w(a)$ for all $w < \infty$.
\end{lemm}
\Pr
The result follows from the following inequalities:
\begin{equation} 
\pi_n^w(a) \le \widehat{\pi}_n^w(a) \le (w+1) \pi_n^w(a) = \exp[\kappa_0^w(a) n + o(n)].
\end{equation}
\qed

\begin{lemm}
When $d \ge 2$, 
$\sup_w \kappa_0^w(a) = \kappa_0(a)$.
\end{lemm}
\Pr 
Clearly $\kappa_0^w(a) \le \kappa_0(a)$ for all $w$.
Write $n=Nr + q$, $0 \le q < N$.  Concatenate $r$ polygons each with $N$ edges, 
and a final polygon with $q$ edges,
using the concatenation construction detailed in Section 4 of \cite{HamWhitt1985}.  
Each polygon with $N$ edges has span no larger than $w=N/2$ so the resulting 
polygon (with $Nr+q$ edges) has span no larger than $w=N/2$.  By the 
argument leading to  (4.14) in \cite{HamWhitt1985} this gives
\begin{equation}
\fl \quad \quad
n^{-1} \log \widehat{\pi}_n^w(a) \ge N^{-1} (1-q/n) \log P_N(a,1) -2 N^{-1} (1-q/N) \log[(d-1) N^{d-1}].
\end{equation}
Hence 
\begin{equation}
\kappa_0^w(a) \ge N^{-1} \log P_N(a,1),
\end{equation}
where we recall that $w = N/2$.  As $N \to \infty$ the right hand side goes to $\kappa_0(a)$
so $\sup_w \kappa_0^w(a) \ge \kappa_0(a)$ which completes the proof.
\qed

This allows us to prove the following:

\begin{theo}
When $d \ge 2$ 
$$\psi_0(a,y) =  \psi_0(a,1) = \kappa_0(a)$$
for all $0 < y \le 1$.
\end{theo}
\Pr 
Fix $0 < y \le 1$.
By monotonicity
$P_n(a,y) \le P_n(a,1)$
so
\begin{equation}
\limsup_{n\to\infty} n^{-1} \log P_n(a,y) \le \kappa_0(a).
\label{eqn:limsupslab}
\end{equation}
By considering one term in the partition function
\begin{equation}
P_n(a,y) \ge y^w \sum_v p_n(v,w)a^v = \pi_n^w(a),
\end{equation}
and 
\begin{equation}
\liminf_{n\to\infty} n^{-1} \log P_n(a,y) \ge \lim_{n\to\infty} n^{-1} \log \pi_n^w(a) = \kappa_0^w(a)
\end{equation} 
for all $w > 0$.  Hence
\begin{equation}
\liminf_{n\to\infty} n^{-1} \log P_n(a,y) \ge \sup_w \kappa_0^w(a) = \kappa_0(a)
\label{eqn:liminfslab}
\end{equation}
Then (\ref{eqn:limsupslab}) and (\ref{eqn:liminfslab}) prove the theorem.
\qed

In particular, this proves that the phase boundary between the free and adsorbed phases
is a vertical line in the $(a,y)$-phase diagram, at $a=a_c^0$.

\section{The phase diagram}
\label{sec:phasediagram}

In this section we give a brief summary of what is known rigorously about the 
form of the phase diagram in the $(a,y)$-plane.

There is a free phase when $a < a_c^0$ and $y < 1$.  When $a > a_c^0$ and $y < 1$ the system
is in an adsorbed phase and $\psi^0(a,y) = \kappa^0(a)$, independent of $y$.

Suppose that $y_I(a)$ is the solution of the equation $\lambda_0(y)=\kappa(a)$ and 
suppose that $y_{II}(a)$ is the solution of the equation $\lambda_0(y) = \kappa_0(a)$.
When $d \ge 3$ $\kappa_0(a) = \kappa(a)$ so $y_I(a) =y_{II}(a)$.
For this case ($d \ge 3$) there is a ballistic phase (where $\psi_0(a,y) = \lambda_0(y)$)
when $y > \max[1,y_I(a)]$ and an adsorbed
phase (where $\psi_0(a,y) = \kappa_0(a)$) when $a > a_c^0$ and $y < y_I(a)$.

When $d=2$ we know less.  The system is in a ballistic phase when $y > \max[1,y_I(a)]$ but we 
do not know whether $y = y_I(a)$ is a boundary of this phase.  When $y < y_{II}(a)$ the system is 
no longer ballistic.  There are two possibilities:
\begin{enumerate}
\item
The phase boundary of the ballistic phase, $y_B(a)$, is equal to $y_{II}(a)$.  In this case there are three
phases: a free phase, a ballistic phase and an adsorbed phase (when $a>a_c^0$ and $y < y_{II}(a)$).
\item
The phase boundary of the ballistic phase satisfies $y_{II}(a) < y_B(a) \le y_{I}(a)$.  Then there at
least four phases: a free phase, a ballistic phase, an adsorbed phase (including the region defined
by $a > a_c^0$ and $y<1$) and at least one additional phase where the free energy depends on both
$a$ and $y$, in the region defined by $a > a_c^0$ and $1 < y < y_B(a)$.  We do not know if $y=1$ is a 
phase boundary.
\end{enumerate}

\begin{figure}[t]
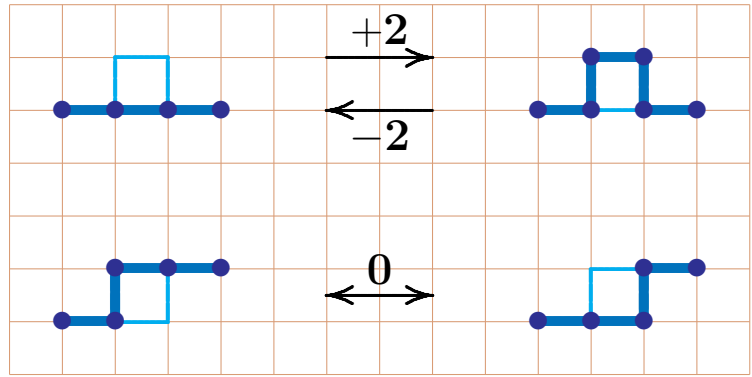

\input figureBFACF.tex
\caption{BFACF moves in the square lattice \cite{BF81}.  A positive move (or a
\textit{positive BFACF move}) increases the length of a polygon by
2 steps by replacing an edge with three edges.  The reverse of this move
reduces the length of the polygon by 2 steps and is a 
\textit{negative BFACF move}.  \textit{Neutral BFACF moves}
change the polygon locally as shown without changing its length.}
\label{figureBFACF}
\end{figure}

\section{Numerical results from Monte Carlo data}
\label{sec:numericalMC}

In order to investigate the phase behaviour in the half square lattice, we collected
approximate enumeration data for polygons as a function of $a$ and $y$
using the GARM algorithm \cite{RJvR08}.  This algorithm was implemented 
using BFACF-style elementary moves \cite{BF81} (see figure \ref{figureBFACF})
on polygons in the half-lattice.  

Polygons in the upper half square lattice and constrained to have at least one edge in
the boundary of the half-lattice (the $x$-axis) can be sampled by executing BFACF 
moves.  This is done as follows using a GARM implementation of BFACF moves.

A positive BFACF move can be done along the polygon 
by replacing one edge by three
\beginpicture 
\setcoordinatesystem units <0.75pt,0.75pt>
\plot 0 0 10 0 /
\arrow <5pt> [.2,.67] from 15 5 to 30 5 
\plot 40 0 40 10 50 10 50 0 /
\multiput {\Large.} at 0 0 10 0 40 0 40 10 50 10 50 0 /
\put {$ $} at 58 0
\endpicture
while maintaining the constraints that the polygon has to step at least
once in the $x$-axis and must stay in the upper half square lattice.  The collection 
of all possible positive BFACF moves is the \textit{positive atmosphere} of the polygon.
The number of possible BFACF moves in the positive atmosphere of a polygon $\omega$
is denoted $a_+(\omega)$.

A negative BFACF move is the reversal of a positive BFACF move, and is
implemented by replacing three edges by one: 
\beginpicture 
\setcoordinatesystem units <0.75pt,0.75pt>
\plot 0 0 0 10 10 10 10 0 /
\arrow <5pt> [.2,.67] from 15 5 to 30 5 
\plot 40 0  50 0 /
\multiput {\Large.} at 0 0 0 10 10 10 10 0 40 0 50 0 /
\put {$ $} at 58 0
\endpicture,
while maintaining the constraints that the polygon has to contain an edge in the
$x$-axis and stay in the upper half square lattice. The collection of all possible 
negative BFACF moves is the \textit{negative atmosphere} of the polygon.
The number of possible BFACF moves in the negative atmosphere of a
polygon $\omega$ is denoted $a_-(\omega)$.

Neutral BFACF moves are implemented by local changes involving
two edges of the polygon: 
\beginpicture 
\setcoordinatesystem units <0.75pt,0.75pt>
\plot 0 0 0 10 10 10 /
\arrow <5pt> [.2,.67] from 15 5 to 40 5 
\arrow <5pt> [.2,.67] from 40 5 to 15 5 
\plot 50 0  60 0 60 10 /
\multiput {\Large.} at 0 0 0 10 10 10 50 0 60 0 60 10  /
\put {$ $} at 68 0
\endpicture
while maintaining the constraints that the polygon has to contain an edge in the
$x$-axis and stay in the upper half square lattice. The collection of all possible 
neutral BFACF moves is the \textit{neutral atmosphere} of the polygon.
The number of possible BFACF moves in the neutral atmosphere of a 
polygon $\omega$ is denoted $a_0 (\omega)$.

A sequence $\phi = \langle \phi_0, \phi_1 , \phi_2 , \ldots, \phi_n \rangle$ 
of polygons in the half square lattice can be sampled by executing a move
uniformly selected from the positive and neutral atmospheres of 
$\phi_j$ to find $\phi_{j+1}$.  The sequence is started in the polygon $\phi_0$ of
length $4$ with one edge in the $x$-axis.  Notice that any polygon 
$\phi_j$ has at least one atmospheric move which can be executed on it,
and that every polygon $\omega$ can be obtained in the upper half
square lattice from $\phi_0$ by executing positive and neutral BFACF
moves.  (To see this, reverse the  steps by starting at $\omega = \phi_n$, 
and show that it can be made shorter by executing a negative atmospheric
move, sometimes after a neutral move was done a number of times).

Let $\phi_n$ be a state with $v$ visits and top plane of height $h$.  The 
probability of generating a sequence $\phi$, starting at $\phi_0$ and
ending in $\phi_n$, is given by
\begin{equation}
P(\phi) = \prod_{j=0}^{n-1} \frac{1}{a_0(\phi_j) + a_+(\phi_j)} .
\end{equation}
Assign a weight
\begin{equation}
W(\phi) = \prod_{j=0}^{n-1} \frac{a_0(\phi_j) + a_+(\phi_j)}{a_0(\phi_{j+1}) + a_-(\phi_{j+1})}
\end{equation}
to the squence $\phi$.

The \textit{average weight} of sequences ending in the state $\phi_n$ is given by
\begin{equation}
\langle W(\phi_n) \rangle = \sum_{\phi \to \phi_n} P(\phi) \, W(\phi) 
= \sum_{\phi \to \phi_n} \prod_{j=1}^{n} \frac{1}{a_0(\phi_j) + a_-(\phi_j)} .
\end{equation}
This, however, is the probability of the \textit{reverse sequence} starting in the state
$\phi_n$ and ending in the state $\phi_0$ if only negative and neutral moves are
done.  This probability is equal to $1$, 
since these sequences end up in state $\phi_0$ with probability $1$.  In other words,
\begin{equation}
\langle W(\phi_n) \rangle  
= \sum_{\phi \to \phi_n} \prod_{j=1}^{n} \frac{1}{a_0(\phi_j) + a_-(\phi_j)}  = 1 .
\end{equation}
This is the GARM counting theorem \cite{RJvR08}.  

If $S_n(v,w)$ is the set of all polygons of length $n$ in the half-lattice with $v$ visits
and height $w$, then the average weight of sequences ending up in states
in $S_n(v,w)$ is given by 
\begin{eqnarray}
W_{v,w} &=& \sum_{\phi_n\in S(v,w)} \langle W(\phi_n) \rangle  \cr
&=& \sum_{\phi_n\in S(v,w)}\sum_{\phi \to \phi_n} \prod_{j=1}^{n} \frac{1}{a_0(\phi_j) + a_-(\phi_j)}  = 
\sum_{\phi_n\in S(v,h)}1  = p_n(v,w).
\end{eqnarray}
In other words, by computing the average weight $W_{v,w}$ of polygons of length $n$
with $v$ visits and height $w$, estimates of the microcanonical partition function
$p_n(v,w)$ are obtained.  This is an example of approximate enumeration
\cite{Rensburg2010} and these data can be used to determine average number of visits or height
for polygons of fixed length.  

The algorithm is implemented with pruning and enrichment in exactly the same
way the PERM or flatPERM algorithms are implemented.  For details, see
references \cite{Grassberger1997} and \cite{Prellberg2004}.  The resulting implementation
of \textit{flatGARM} is a flat histogram sampling method which continually 
prunes states of low weight that do not contribute much to the partition 
function, and otherwise enriches states of high weight in the sampling.  The algorithm was run to complete
about $11,000$ GARM sequences and we collected data for polygons with up to
$200$ edges and computed the free energy, the mean number of visits and 
the mean height of the polygons, as well as the variances of these quantities.

\begin{figure}[th!]
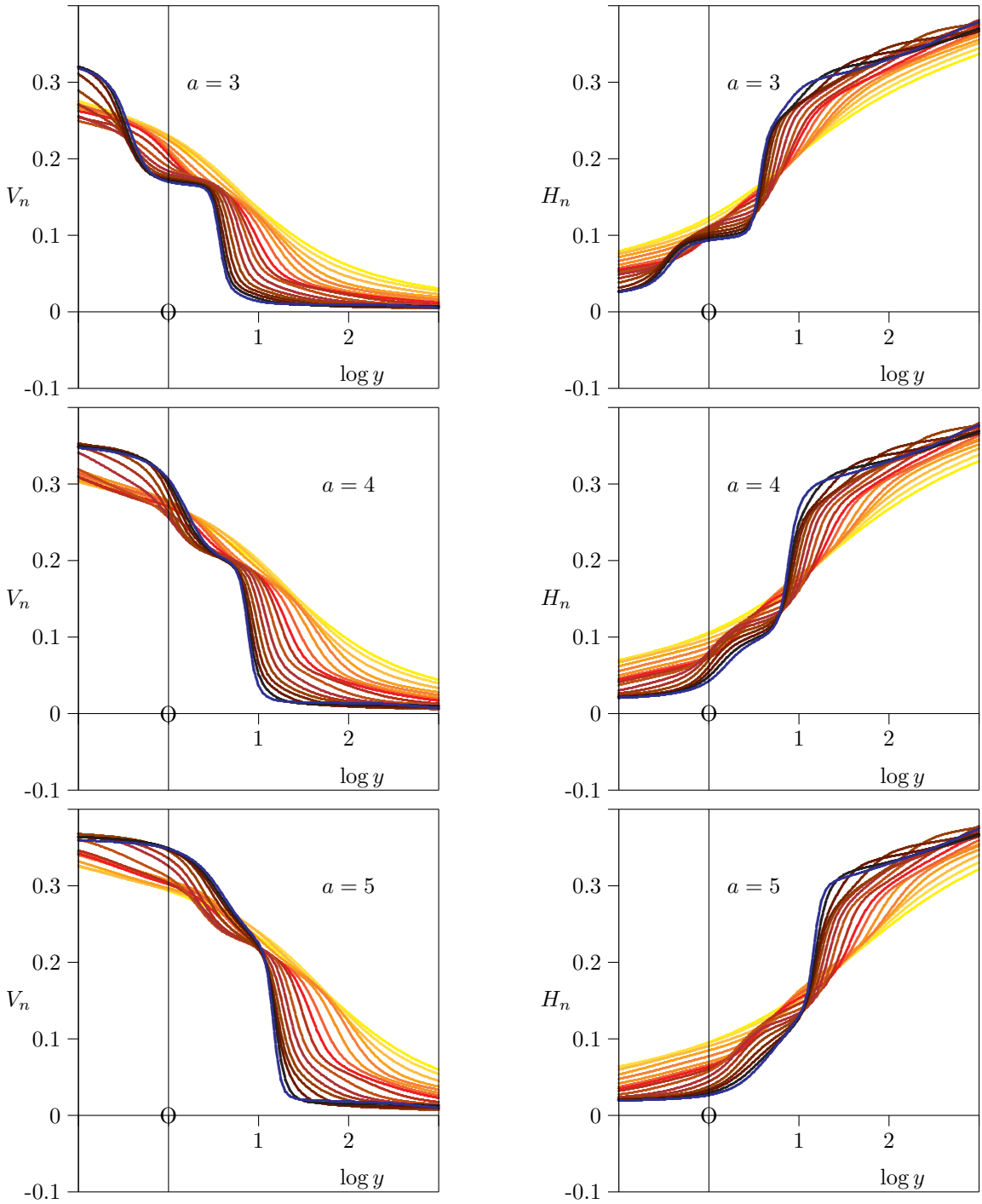

\input figureEa-N.tex
\caption{Energies of pulled adsorbing polygons.
Left panels: The average density of visits $V_n$ as a function of $\log y$ for $a=3$
(top left panel), $a=4$ (middle left panel) and $a=5$ (bottom left panel).
Right panels: The average height $H_n$ as a function of $\log y$ for $a=3$
(top right panel), $a=4$ (middle right panel) and $a=5$ (bottom right panel).
In all these graphs the value of $n$ increased from $40$ to $200$
in steps of $10$, with curves progressively darker as $n$ increases.}
\label{figureEa}  
\end{figure}

\begin{figure}[t]
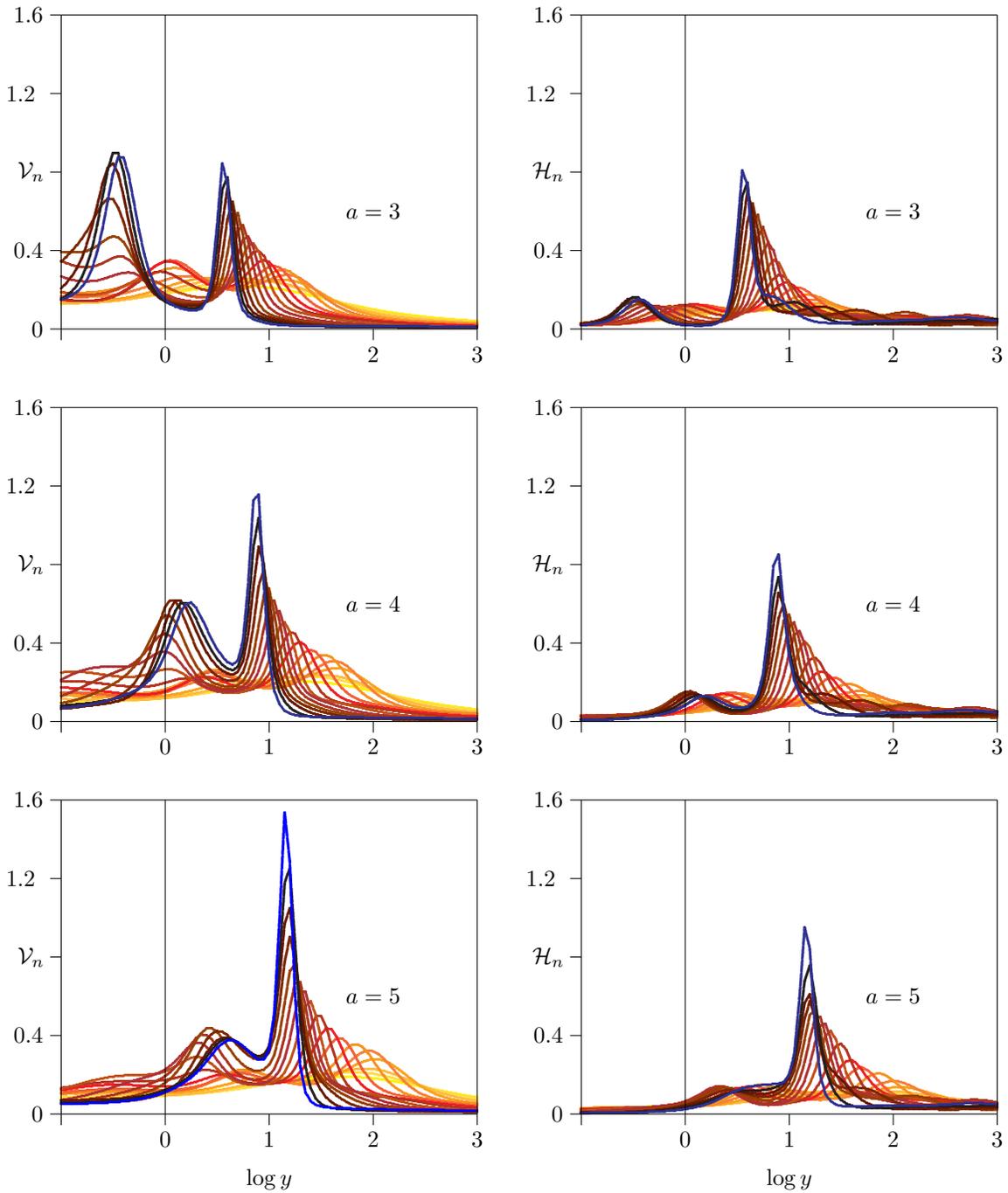

\input figureCa-N.tex
\caption{Variances of pulled adsorbing polygons.
Left panels: The variance $\C{V}_n$ of the density of visits as a function of $\log y$ for $a=3$
(top left panel), $a=4$ (middle left panel) and $a=5$ (bottom left panel).
Right panels: The variance $\C{H}_n$ of the scaled height as a function of $\log y$ for $a=3$
(top right panel), $a=4$ (middle right panel) and $a=5$ (bottom right panel).
In all these graphs the value of $n$ increased from $40$ to $200$
in steps of $10$, with curves progressively darker as $n$ increases.}
\label{figureCa}  
\end{figure}

In Figure \ref{figureEa} we show the average value of the density of visits 
$V_n$ and the average value of the scaled height $H_n$ plotted against 
$\log y$ for $a=3$, $a=4$ and $a=5$.  There is clear
evidence for two phase transitions in these figures.  We know that for $\log y < 0$
the density of visits and scaled height are independent of $y$.  In all four
panels the curves seem to be approaching horizontal lines as $n$ increases in this
regime.  There is a transition to a phase with a reduced number of visits and a larger
value of the height around $y=1$, and a second, more marked, transition at  a location
whose value depends on $a$ from this phase to a ballistic phase where the density of
visits approaches zero.  In the intermediate regime between the two transitions the
density of visits and the average scaled height depend on both $a$ and $y$.  This regime
is referred to as a \textit{mixed phase}.

For instance, when $a = 5$, the curves for $V_n$ are approaching a horizontal line
when $\log y \leq  0$, as we know must happen from our rigorous arguments. For 
$\log y > 0$ there is a regime in which $V_n$ is a decreasing function of $\log y$ and 
$H_n$ is an increasing function of $\log y$, so the free energy depends on $y$. 
It also depends on $a$, as can be seen by comparing with the results at 
$a=3$ and $a=4$.
When $\log y$ is somewhat larger than $1$ there is a rapid decrease in $V_n$ and a rapid increase
in $H_n$ and then at larger values of $y$ both quantities become less dependent on $y$.
All of this suggests two transitions, one from an adsorbed phase where the free
energy only depends on $a$ to a mixed phase where the free energy depends on both
$a$ and $y$, and a second transition at larger $y$ to a ballistic phase. In this third
(ballistic) phase the free energy depends on $y$ but is essentially independent of $a$.
We know rigorously that in the infinite $n$ limit the free energy is independent of $a$
in the ballistic phase.

In Figure \ref{figureCa} we show the corresponding fluctuation quantities $\C{V}_n$ and
$\C{H}_n$ of the data in
Figure \ref{figureEa}.  In all four panels there are two peaks consistent with two phase
transitions.  If we look first at the transitions at larger values of $y$ then the peaks
are growing and moving to the left with increasing $n$.  This is clear evidence of
a phase transition although it is difficult to determine a precise location for the
transition.  The peaks at smaller values of $y$ are more difficult to interpret.  At
$a=3$ the peaks largely occur for $y<1$ but the positions may be moving towards
$y=1$ with increasing $n$.  However, we know rigorously that there is no transition
for $y<1$.  When $a=5$ the picture is clearer.  The peaks occur for values of
$y>1$ though their positions do not move smoothly with increasing $n$.

We interpret these results as showing the existence of a mixed phase in 
this model.  There is a transition from the mixed phase to the ballistic phase occurring
at a  critical value of $y$ that is a function of $a$.  There is also a second transition
from the adsorbed phase to the mixed phase which must occur at a value of
$y\geq 1$, but we cannot be sure exactly where it occurs, or whether its location is
a function of $a$.

\section{Series Analysis}
\label{sec:numerical}

\subsection{Exact enumerations \label{sec:enum}}

Our algorithm for the enumeration of self-avoiding polygons (SAP) on the square lattice 
is based on the work of Enting \cite{Enting80} who  pioneered the use of the finite lattice method.  
The first terms in the generating  function for SAP are calculated using transfer matrix (TM)
techniques to count  the number of polygons in rectangles $W$ unit cells wide and $L$ cells long. 
Any polygon spanning such a rectangle has a  size of at least $2(W+L)$ edges. Adding contributions 
from all rectangles of width $W \leq W_{\rm max}$ and length $W \leq L \leq 2W_{\rm max}-W+1$  
the number of polygons per vertex of an infinite lattice is obtained correctly up to length 
$N=4W_{\rm max}+2$.  Normally one can use the symmetry of the square lattice to restrict the TM 
calculations to rectangles with  $W\leq W_{\rm max}/2$ and $L\geq W$ by counting contributions for rectangles  
with $L>W$ twice. The interactions with the surface breaks the symmetry and therefore we have to consider all 
rectangles with $W\leq W_{\rm max}$. The size of the transfer matrix grows exponentially with $W$ and to 
partially overcome this hurdle we break the TM calculation on the set of rectangles 
into two sub-sets with $L\geq W$ and $L<W$, respectively. In the calculations for the 
sub-set with $L\geq W$  the surface is placed on the bottom of the rectangle and
for the sub-set with $L< W$  the surface is placed on the left side of the rectangle.
The height (or span) $h$ of the SAP is simply $W$ for the first sub-set and $L$
for the second sub-set.

\begin{figure}[t]
\begin{center}
\includegraphics[width=0.45\textwidth]{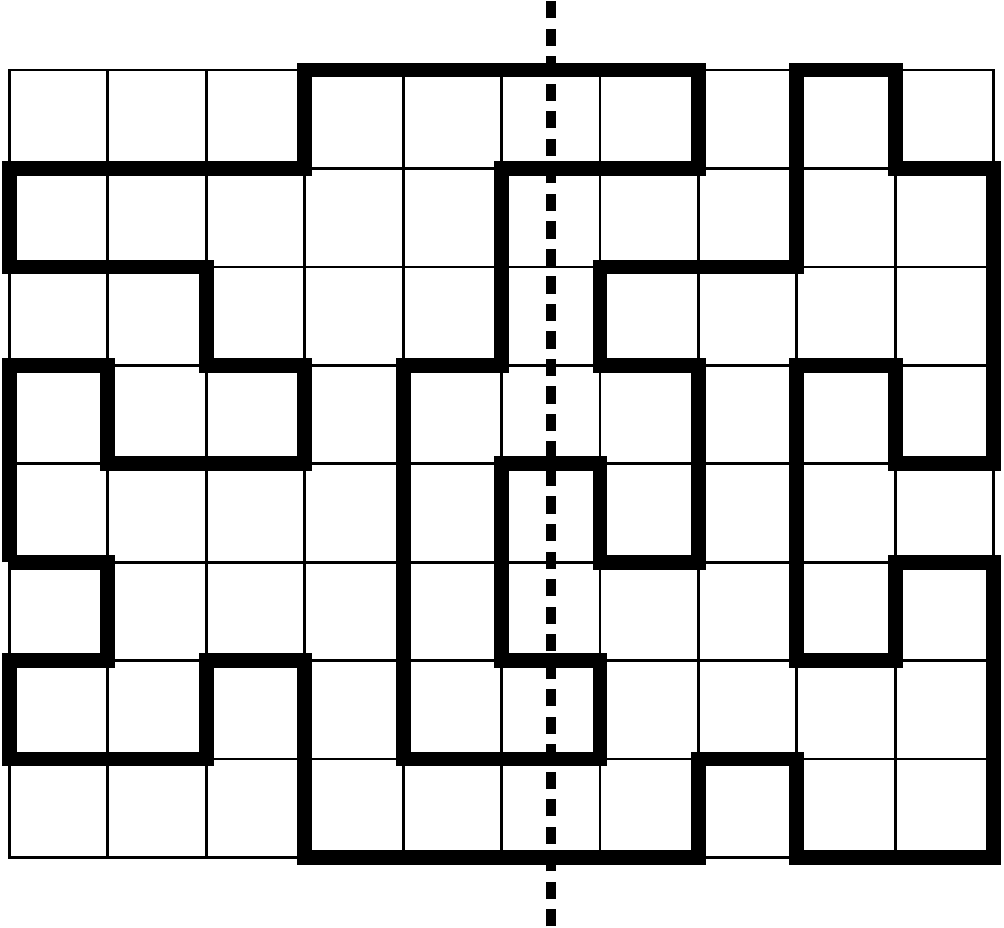}
\hspace{1cm}
\includegraphics[width=0.45\textwidth]{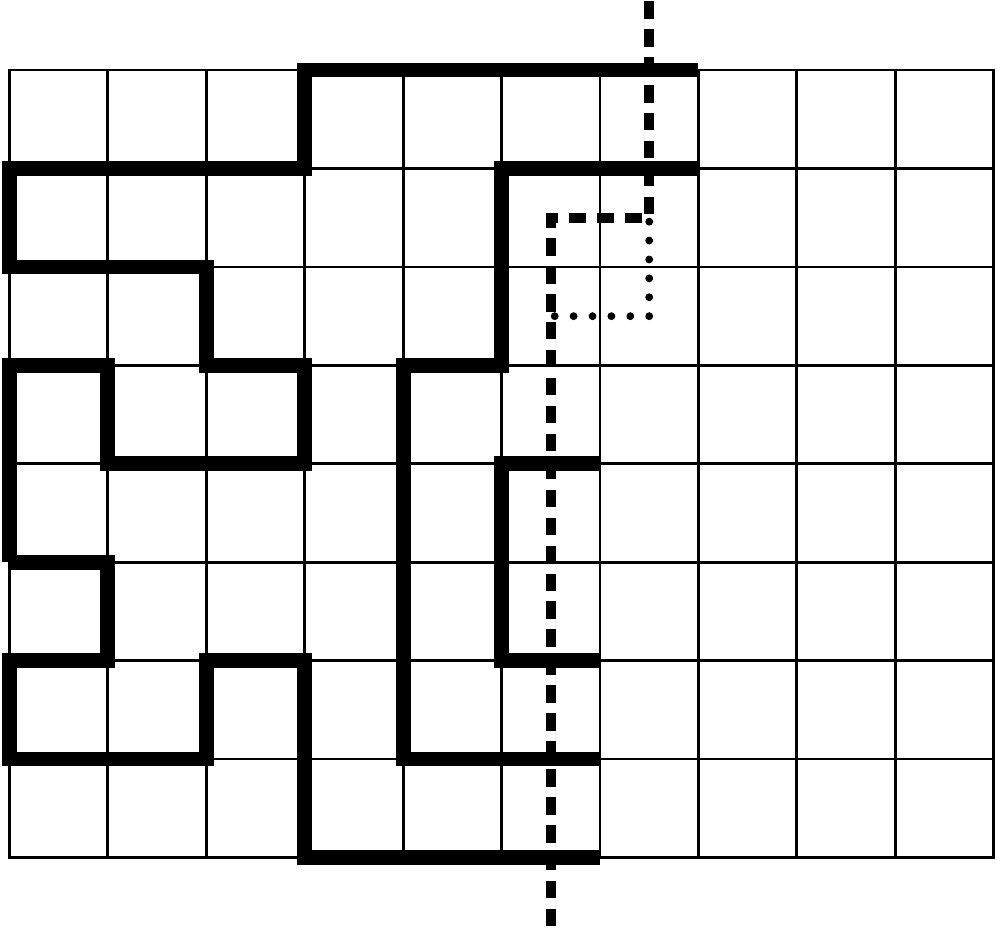}
\end{center}
\caption{
The first panel shows an example of a self-avoiding polygon on a $8\times 10$ rectangle with a surface on
the bottom having 8 vertices in the surface and span 8. Alternatively we can view it as a SAP on a $10\times 8$ rectangle
with the surface on the left  having 7 vertices in the surface and span 10. 
The cut-line (dashed-line) splits the SAP into a set of arcs to the left (right) of the cut-line.
The second panel illustrates how the cut-line is moved in order to build the rectangle vertex by vertex.}
\label{fig:sapex}
\end{figure}

The basic idea of the  algorithm is illustrated by the  example in Figure~\ref{fig:sapex}.
Clearly any SAP is topologically equivalent to a circle and when cut by 
a vertical line  (the dashed line in Figure~\ref{fig:sapex}) it is broken into 
several arcs on {\em either} side of the cut-line  connecting two occupied edges.
As the cut-line is moved from left to right we keep track of the ever changing connections (arcs)
between occupied edges on the cut-line.  Each end of an arc is assigned one of two labels 
depending on whether it is the lower or upper edge. Any configuration along the
cut-line can thus be represented by a set of edge states $\{\sigma_i\}$, where

\begin{equation}\label{eq:states}
\sigma_i = \left\{ \begin{array}{rl}
0 &\;\;\; \mbox{empty edge}, \\ 
1 &\;\;\; \mbox{lower edge}, \\
2 &\;\;\; \mbox{upper edge}. \\
\end{array} \right.
\end{equation}
\noindent
 Since crossings are not permitted this encoding uniquely describes 
how the occupied edges  are connected.  Reading from bottom to top  the configuration or signature $S$ 
along the cut-line of the SAP in Figure~\ref{fig:sapex} is  $S=\{111020022 \}$ encoding the arcs to the left of the cut-line.  
The most efficient implementation of the TM algorithm involves moving 
the cut-line in such a way as to build up the lattice vertex by vertex (see the second panel of Figure~\ref{fig:sapex}). 
The sum over all contributing SAP is calculated as the cut-line is moved through the lattice. 
For each signature we maintain a generating function $G_S$ for partially completed polygons. Here $G_S$ is a truncated  
polynomial  $G_S(x,a)$ where $x$ is conjugate to the number of edges in the partially completed polygon and $a$ 
to the number of visited vertices in the surface.  In a TM update each source signature $S$ (before the boundary is moved) gives rise 
to only a  few new target signatures  $S'$ (after the move of the boundary line).
In a specific update $k=0, 1$ or 2 new edges are occupied and $m=0$ or 1 surface vertices are added 
(on the bottom or left of the rectangle depending on the sub-set we are dealing with) leading to the
update  $G_{S'}(x,a)=G_{S'}(x,a)+x^ka^mG_S(x,a)$.  In the case illustrated in Figure~\ref{fig:sapex} the two `new' edges 
intersected by the dotted lines are either empty ($k=0$) or  occupied  ($k=2$) if a new arc is inserted.

We calculated the number of SAP up to length $N=100$. The calculation was
performed in parallel using up to 32 processors, a maximum of  some 70GB of memory 
and using a total of just under 1000 CPU hours.  Details of the implementation 
and parallelization of our algorithm can be found in \cite{Jensen99,Jensen03,GJ09}.

\subsection{Results}
\label{sec:results}

For SAPs in the bulk, on a bi-partite lattice such as the square, simple-cubic, or indeed hyper-cubic lattice, it is universally believed (though not proved) that
\begin{equation}
p_{2n} \sim const \cdot \mu^{2n} \cdot n^{\alpha_b-3},
\end{equation}
where, for the square lattice, $\alpha_b = 1/2,$ while for the simple-cubic lattice the best estimate \cite{C10} is $\alpha_b \approx 0.237209.$
However if the polygon sits at a surface and a compressive force (i.e. $y < 1$) is applied to the top of the polygon, then it has recently been shown by Beaton {\em et al.} \cite{BGJL15} from probability arguments and particularly assuming SLE predictions that the
expected asymptotics now includes a stretched-exponential term. More precisely, for the square lattice,
\begin{equation}
p_{2n} \sim const \cdot \mu^{2n}\cdot \mu_1^{n^{3/7}} \cdot n^{-11/7},
\end{equation}
where both the constant and $\mu_1$ are $y$-dependent, and the $y$ dependence of $\mu_1$ is also predicted.

In this Section we describe the results from series analysis, chiefly using the method of
differential approximants (DAs) \cite{GJ09}.  Unfortunately, as discussed in \cite{G15}, this method has some problems when applied to generating functions whose coefficients have stretched-exponential terms. In particular the estimate of the dominant growth constant $\mu$ produced is much less precise than is usually the case, while the estimates of the critical exponent vary wildly from approximant to approximant. In practical terms, we expect accuracy of 2-4 significant digits in the critical point estimate, while the critical exponent estimate is unobtainable by this method.

We first discuss the $y$-dependence of the free-energy
$\lambda_0 (y)$ when there is no surface interaction (i.e. $a=1$), then the $a$-dependence of the 
free-energy $\kappa_0 (a)$ when there  is no applied force (i.e., $y=1$) and finally the two variable 
free-energy $\psi_0 (a,y)$ when there is both a surface interaction and an applied force.

\subsubsection{No surface interaction. $a=1.$}\label{sec:lambda}

\begin{figure}
\centering
\includegraphics[scale =0.5] {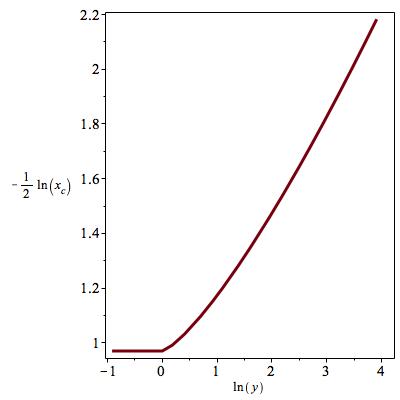}
 \caption{The $y$-dependence of the free-energy $\lambda_0 (y)$. 
}
 \label{fig:lambda}
\end{figure}

If we write 
\begin{equation}
H(x,y) = \sum_n P_{2n}(1,y) x^n=\sum_n e^{2\lambda_0(y) n + o(n)} x^n
\end{equation}
where $x$ is the generating variable conjugate to the half-length of the polygon,
then $H(x,y)$ will be singular at $x=x_c(y) = \exp[-2\lambda_0 (y)]$
and, close to this singularity,  $H(x,y)$ is expected to behave as 
\begin{equation}
H(x,y) \sim {A}\, {[x_c(y)-x]^{\alpha (y)}}
\end{equation}
where $\alpha(y)$ is a critical exponent whose value depends on $y$.

In the last three columns of  Table~\ref{tab:y1}  below we give the results of an analysis of the series  $H(x,y)$ for various values of $y$.
The resulting estimates of the free-energy $\lambda_0 (y) = -{\frac{1}{2}\log x_c}$ are plotted in Figure~\ref{fig:lambda}.
The series were analysed using second and third order differential approximants \cite{GJ09}. At $y=1$ the series is well behaved and has critical point $1/\mu^2$ with exponent $\alpha=3/2,$ the exponent for self-avoiding polygons, which is unchanged if we consider SAPs attached to a surface (unlike the SAW case).

For $y$ just below 1 the series are quite difficult to analyse, due to the presence of the stretched exponential term. Estimates of $x_c$ are moderately close to the known value $ 1/\mu^2$ in magnitude, but just below $y=1$ they have a small imaginary part. As we lower $y$ further below 1, we get approximants moderately close to  $ 1/\mu^2$  with very large exponent values, and poor convergence. This is exactly the behaviour discussed in \cite{G15} when using the method of differential approximants to analyse series with a stretched-exponential term. The data are consistent with $\mu$ fixed at the bulk (no force) value, and indeed we have proved that in the free region the free-energy stays at the bulk value.

For $y \ge 1.5$ the series are beautifully behaved, the singularity is clearly seen to be a square root, and we can provide 10 digit (or more) accuracy in estimates of the critical point.
For $1 < y < 1.5$ we get the sort of behaviour we expect with a discontinuous change in exponent as we transition from an exponent $3/2$ to a square root.

So, in summary, it appears that for $y < 1$ we have $x_c = 1/\mu^2$ and stretched exponential behaviour;  for $y=1$ we have $x_c=1/\mu^2$ and exponent $\alpha = 3/2$ and for $y > 1$ we have $x_c$ monotonically decreasing as $y$ increases, and with a square root singularity.

\subsubsection{No applied force. $y=1.$}\label{sec:kappa}

Define the generating function
\begin{equation}
K(x,a) = \sum_n P_{2n}(a,1) x^n=\sum_n e^{2\kappa_0(a) n + o(n)} x^n.
\end{equation}
$K(x,a)$ will be singular at $x=x_c(a) = \exp[-2\kappa_0 (a)]$
and, close to this singularity,  $K(x,a)$ should behave as 
\begin{equation}
K(x,a) \sim {B}\,{[x_c(a)-x]^{\alpha (a)}}
\end{equation}
where $\alpha (a)$ is a critical exponent whose value depends on $a$.

\begin{figure}
\centering
\includegraphics[scale =0.5] {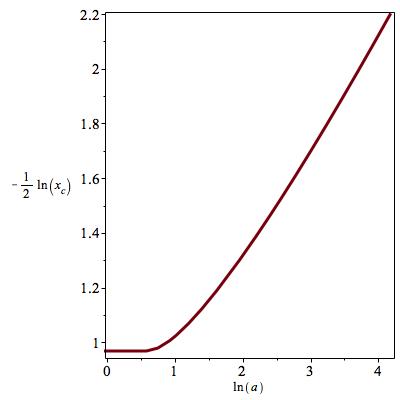}
 \caption{The $a$-dependence of the free-energy $\kappa_0 (a)$.
}
 \label{fig:kappa}
\end{figure}
We have analysed the series  $K(x,a),$ corresponding to the ``no force'' situation, so that $y=1.$ Here the differential approximants work well, as there are no stretched-exponential terms and the critical point and exponent are well estimated.  The results are shown in the first three columns of Table~\ref{tab:y1}. If we denote the transition
 from the free phase to the adsorbed phase by $a_c^o=\exp(-\epsilon/k_B T_c^o)$, and denote the corresponding quantity for adsorbed self-avoiding walks
by $a_c=\exp(-\epsilon/k_B T_c),$ we have proved that $a_c \le a_c^o.$ The numerical evidence is extremely strong that equality holds. 
 The best estimate \cite{Guttmann2014} for the SAW case is $a_c = 1.775615 \pm 0.000005.$   From  Table \ref{tab:y1}, we see that at this value of $a$ the exponent is estimated to be $0.754$ which is reasonably close to the conjectured exact value \cite{Duplantier1990}  $\alpha^{sp}=3/4,$ where the superscript refers to the ``special'' transition that takes place right at the adsorption temperature. 
 
 Note that for $a < a_c$ the exponent is 3/2. At $a_c$ it has changed (presumably discontinuously) to $3/4.$ We looked at nearby values, and found that at $a=1.774$ the exponent appeared to be $0.763$ reflecting a cross-over from 1.5 to 0.75, while at $a=1.776$ the free-energy has started to change, as the estimate of $x_c$ was 0.1436799, while the exponent was around $0.751.$ Thus from the exponent value at $a=1.774$ and the free-energy value at $1.776$ we conclude that $a_c$ lies between these two values. It therefore seems very likely that $a_c^o=a_c,$ and if not, they differ by less than 1 part in a thousand, which seems very unlikely.
 
 So in summary it seems that for $a=a_c$ the singularity is characterised by an exponent $3/4,$ and that this changes discontinuously to a square root for $a > a_c.$ For $a < a_c$ the exponent is, as we would expect, given by $\alpha=3/2.$  
In Figure~\ref{fig:kappa}  we give our estimates of the free-energy $\kappa_0 (a) = -{\frac{1}{2}}\log x_c$ as a function of $\log a$.

\begin{table}
   \centering
   \begin{tabular}{|l|l|l||l|l|l|}
   \hline
      $a \,\,(y=1)$    & $x_c$ & Exponent& $y \,\,(a=1)$& $x_c$ & Exponent\\
\hline
0.5 & 0.143680629 & 1.5000 & 0.4 & 0.147 & 20 \\
1 & 0.143680629  &1.5000  & 0.7 & 0.1456  &11 \\
1.775385 & 0.143680629 & 0.754 &0.9 & 0.1432 $\pm$ 0.0004$i$  & complex \\
 2.1 & 0.1406445 & 0.5000 & 1 & 0.143680629  &1.5000 \\
2.5 & 0.1332540 & 0.5000 & 1.2 & 0.1377 & 0.6 \\
2.75 & 0.1282078 & 0.5000 & 1.5 & 0.12702 & 0.495\\
3.3 & 0.1175624 & 0.5000 & 2 & 0.1118410 & 0.4998 \\
4 & 0.1058177 & 0.5000 & 2.5 & 0.1000544& 0.5000\\
5 & 0.09243473 & 0.5000 & 3 & 0.09075811& 0.5000\\
7 & 0.07390853 & 0.5000 & 4 & 0.07703333& 0.5000\\
9 & 0.06179279 & 0.5000 & 5 & 0.06733037& 0.5000\\
11 & 0.05324182 & 0.5000 & 7 & 0.05436885& 0.5000\\
13 & 0.04686712 & 0.5000 & 9 & 0.04598874& 0.5000\\
16 & 0.03983702 & 0.5000 & 12 & 0.03768982& 0.5000\\
20 & 0.03330407 & 0.5000 & 15 & 0.03213498& 0.5000\\
25 & 0.02772693 & 0.5000 & 19 & 0.02701707& 0.5000\\
32 & 0.02253994 & 0.5000 & 25 & 0.02196582& 0.5000\\
40 & 0.01862508 & 0.5000 & 32 & 0.01814603& 0.5000\\
50 & 0.01534307 & 0.5000& 40 & 0.01521102& 0.5000\\
65 & 0.01217273 & 0.5000& 50 & 0.01270767& 0.5000\\
        
  \hline    
   \end{tabular}
 \caption{SAPs at a surface. Estimates of $x_c$ and exponents for $y=1$  and various $a$ values and estimates of $x_c$ and exponents for $a=1$  and various $y$ values.}
   \label{tab:y1}
\end{table}

\subsubsection{The region $y > 1,$ $a < a_c.$}
In this region (and indeed for larger values of $a,$ the precise limit depending on the value of $y$), we are in the ballistic regime. For fixed $y,$ we expect the free-energy to be independent of $a,$ until we cross a phase boundary. That this is the case is shown in the first three columns of Table ~\ref{tab:y2}, where we show the results for $y=5.$ As the value of $a$ increases, the free-energy remains constant until, for $a$ sufficiently large it starts to change with $a.$ This constancy is the expected behaviour in the ballistic regime,
and it is clear that already at $a=4$ we have transitioned to another regime. We give a second example in Table~\ref{tab:y3} where we show data for $y=2.$ Here we see a transition occuring around $a=2.5.$ We examine the nature of this regime below.

\subsubsection{The region $y < 1$ and $a > a_c.$}
We have seen that for $y < 1$ and $a < a_c$ we are in the so-called ``free'' region, the free-energy is constant, but one has stretched-exponential behaviour. For SAWs, when $y < 1$ and $a> a_c$ one is in the adsorbed regime, and that 
is the case also for polygons. However in this regime we still observe stretched exponential behaviour, and a free-energy that depends only on the value of $a,$ and agrees with the value given in Table~\ref{tab:y1} for $y=1,$ though our estimates of the free-energy in this regime are less precise than elsewhere because of the stretched exponential behaviour. For this reason, we are also unable to estimate the associated critical exponent in this region. At $y=1$ for $a=a_c$ the generating function has a square-root singularity. So this is an adsorbed regime, but with a phase boundary at $y=1,$ the nature of which we now examine.

\subsubsection{The region $y > 1$ and $a > a_c.$}
If one chooses a value of $a > a_c,$ then as $y$ increases above 1, we find the free-energy changes monotonically with $y.$ An example of this is shown in Table~\ref{tab:y2}, where in the last three columns we show the results of our analysis with $a=3.45,$ which is about double the value of $a_c.$ As $ y$ increases the estimates of $x_c$ decrease, and though they are not as stable as we might like, the exponent values are initially around $-1,$ suggesting a simple pole. As $y$ gets sufficiently large, the exponents switch to a square-root. What has happened is that we have gone from a mixed regime where the free-energy depends on both $y$ and $a$, to the ballistic regime.

\begin{table}
   \centering
   \begin{tabular}{|l|l|l||l|l|l|}
   \hline
      $a \,\,(y=5)$    & $x_c$ & Exponent& $y \,\,(a=3.45)$& $x_c$ & Exponent\\
\hline
1 & 0.067330372 & 0.5000 & 1.0 & 0.11148643447 & 0.500000 \\
2.5 & 0.06733 & 0.4999  & 1.25 & 0.11  &-1.3 \\
2.75 & 0.06733 & 0.498 & 1.5 & 0.11078& -0.83 \\
3.0 & 0.067332 & 0.477 & 1.75 & 0.1073 & -0.8 \\
3.25 & 0.06734 & 0.3 & 3.5 & 0.0824875 & -1.1\\
3.35 & 0.06734 & 0.1 & 4 & 0.076809 & -1.06\\
3.4 & 0.067343 & -0.1 & 4.5 & 0.07177& -0.73\\
 3.5 & 0.06733 & -0.452 & 5 & 0.06734  &-0.27 \\
4.0 & 0.066687 & -1.08 & 5.5 & 0.063464 & 0.13 \\
5.0 & 0.0632077 & -1.0000 & 6 & 0.060063 & 0.34\\
8.0 &0.05137318 & -1.00000 & 7 & 0.054369 & 0.48 \\

 \hline    
   \end{tabular}
 \caption{SAPs at a surface. Estimates of $x_c$ and exponents for $y=5$  and various $a$ values and estimates of $x_c$ and exponents for $a=3.45$  and various $y$ values.}
   \label{tab:y2}
\end{table}

\subsubsection{Phase diagram calculation}

For SAWs, we were able to locate the phase boundary between the ballistic and the adsorbed region by solving $\kappa(a)=\lambda(y)$ \cite{Guttmann2014}. But for SAPs in  the mixed region the free energy, $\psi_0$, depends on both $a$ and $y,$ so we cannot locate the phase boundary in this way. We know that the phase boundary is on or between the solutions of $\kappa(a)=\lambda(\sqrt{y})$ and $\kappa_0(a)=\lambda(\sqrt{y}).$ As there is a mixed phase, it follows that the boundary of the ballistic phase cannot coincide with the latter solution, and must be at strictly smaller values of $a$ for each $y > 1.$

In fact it seems from our numerical data that the phase boundary does indeed lie on the solution of $\kappa(a)=\lambda(\sqrt{y}).$ Consider the data in Table \ref{tab:y2}. In the first three columns, the data for $y=5$ are given.  As $a$ increases, both the free-energy and exponent initially remain essentially constant, as we expect in the ballistic regime. Then between $a=3.25$ and $a=3.5$ the exponent has changed dramatically, from $0.3$ to $-0.45,$ reflecting, we suggest, the transition from the square-root singularity characteristic of the ballistic regime to the simple-pole behaviour characteristic of the mixed regime. So the phase boundary should lie between these values of $a.$ The mid-point is $a=3.375.$ From the phase boundary for SAWs, given in \cite{Guttmann2014}, we find at $y=\sqrt{5}$ that the point on the phase boundary is at $a=3.379,$ remarkably close to our crude estimate.

\begin{table}
   \centering
   \begin{tabular}{|l|l|l||}
   \hline
      $a \,\,(y=2)$    & $x_c$ & Exponent\\
\hline
1.0 & 0.111841 & 0.4998  \\
1.5 & 0.111841  &0.4996   \\
2.0 & 0.111841 & 0.49  \\
2.5 & 0.111807 & -0.56 \\
2.6 & 0.11157 & -0.94\\
3.0& 0.108584 & -1.01 \\

 \hline    
   \end{tabular}
 \caption{SAPs at a surface. Estimates of $x_c$ and exponents for $y=2$  and various $a$ values, showing exponent change as one crosses the phase boundary.}
   \label{tab:y3}
\end{table}
 
 Now consider the data in the last three columns of Table \ref{tab:y2}. Here $a=3.45,$ and as the value of $y$ increases it is clear that there is a transition from the simple pole behavior for $y \le 4.5,$ toward the square-root behaviour of the ballistic regime when $y > 5.5.$ So we expect the phase boundary to be at around $y=5.$ Again from the phase boundary for SAWs, given in \cite{Guttmann2014}, we find at $a=3.45$ that the point on the phase boundary is at $y=2.2593.$ Squaring this, we expect the corresponding point on the SAP phase boundary to be at $y \approx 5.10,$ again close to our observed transition point.
 
Our third such calculation involves the data in Table \ref{tab:y3}. Here $y=2,$ and it is clear that the transition from the ballistic to the mixed regime takes place at a value of $a$ around 2.5. Turning to the SAW phase boundary, we find that at $y=\sqrt{2},$ the point on the phase boundary is at $a=2.498.$

\begin{figure}
\centering
\includegraphics[scale =0.7] {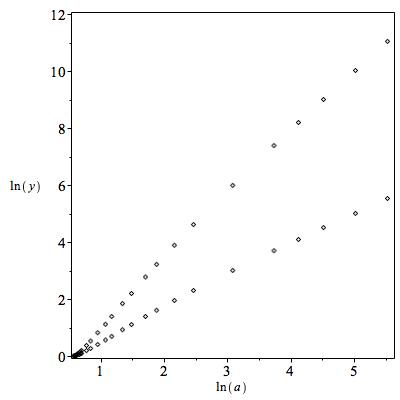}
 \caption{The conjectured phase boundary between the mixed and ballistic phases in the 
 $(\log{a},\log{y})$-plane for SAPs (upper point-plot) and between the adsorbed and ballistic phases for SAWs (lower point-plot). }
 \label{fig:phaseboundary2}
\end{figure}

So while we cannot identify the phase boundary with the precision that was achieved in the SAW case, all the evidence is consistent with the hypothesis that the phase boundary between the ballistic and mixed phases is given by the solution of $\kappa(a)=\lambda(\sqrt{y}).$ Taking this as our working assumption, we show in Figure \ref{fig:phaseboundary2} the phase boundary 
between the mixed and ballistic phases for SAPs (upper point-plot) and the phase boundary between the ballistic and adsorbed phases for SAWs in the lower point-plot, in the $(\log{a},\log{y})$-plane.


\begin{figure}[htbp]
   \centering 
 \includegraphics[scale=0.5]{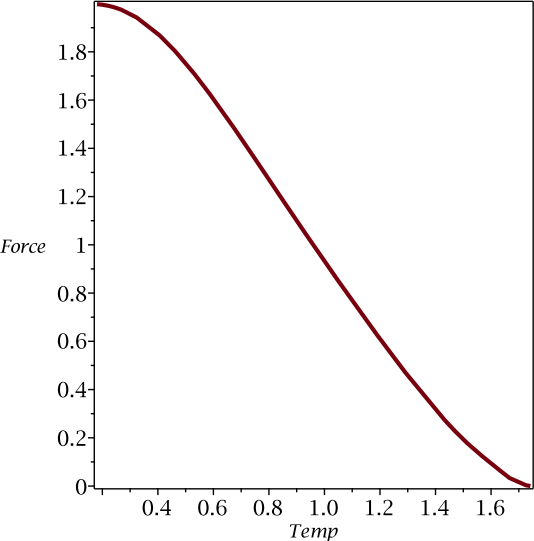} 
\caption{The phase boundary given as a force-temperature diagram. 
The horizontal axis is the temperature $T=\frac{1}{\log(a)}$, the vertical axis is 
the force, given by $f=\frac{\log(y)}{\log(a)}.$}
   \label{fig:ft1}
\end{figure}

We can switch to physical variables (force and temperature)
using equation (\ref{eqn:fugacities}).  Without much loss of generality 
we can set $\epsilon = -1$ and work in units where $k_B=1$.  The corresponding phase
boundary in the force-temperature plane is given in Figure~\ref{fig:ft1}.  Notice that
the force at zero $T$ is $2$ and the limiting slope at $T=0$ is zero.   For the
self-avoiding walk the zero derivative at $T=0$ was predicted in reference 
\cite{JvRW2013}.  The curve is monotone decreasing as $T$ increases, with no re-entrance.

\section{Conclusions}
\label{sec:discussion}

When a polymer is adsorbed at a surface it can be desorbed by applying a 
force normal to the surface to pull the polymer away from the surface.  We
already have a number of rigorous results available for the self-avoiding walk
model of a linear 
polymer \cite{JvRW2013,RensburgWhittington2016b,RensburgWhittington2017}.  
The behaviour might depend on polymer 
architecture and we begin to investigate this issue in this paper by considering 
a self-avoiding polygon model of a ring polymer.

When the dimension is $d \ge 3$ we show that the critical force-temperature curve
(\emph{i.e.} the temperature dependence of the force required to desorb the 
polygon) can be characterized in terms of the free energy of the adsorbed
polygon without a force and the free energy of the polygon subject to a force but not 
interacting with the surface.  Similar results are known for 
the self-avoiding walk model \cite{JvRW2013}, though we also show that the critical
force-temperature curve is different for the polygon case.  We are 
able to determine the phase boundaries in the phase diagram in terms 
of these free energies.  When $d=2$ the situation is more complicated 
because at most half of the vertices of the polygon can be in contact
with the surface.  For $d=2$ we have bounds on the free energy but
our results are less complete.  

Our Monte Carlo and exact enumeration results suggest the existence
of a \emph{mixed phase} in two dimensions where the free energy
depends on both $a$ and $y$. 
The results on adsorbing and pulled staircase polygons in reference \cite{Beaton2017} 
similarly show a mixed phase which is adsorbed-ballistic in the phase 
diagram.  Mixed phases were also seen in a directed model of copolymer
adsorption in reference \cite{IJvR12}.  
The model of adsorbing and pulled staircase polygons  is a directed
version of our model of two dimensional adsorbing 
and pulled polygons.

\vspace{5mm}

\section*{Acknowledgements}
EJJvR and SGW acknowledge support in the form of Discovery Grants from NSERC (Canada).
SGW was partially supported by the Leverhulme Trust Research Programme Grant 
No. RP2013-K-009, SPOCK: Scientific Properties of Complex Knots.  He would like to 
acknowledge the hospitality of University of Bristol where some of this research was carried out.
AJG and IJ acknowledge support in the form of a Discovery Grant DP140101110 from the 
ARC (Australia). The computational work of IJ was 
undertaken with the assistance of resources and services from the National 
Computational Infrastructure (NCI), which is supported by the Australian Government.

\end{document}